\documentclass[preprint]{elsarticle}
\usepackage[english]{babel}
\usepackage{natbib}
\usepackage{hyperref}
\usepackage{lscape}
\usepackage{xcolor}
\usepackage{color}
\usepackage{colortbl}
\usepackage{setspace}
\usepackage{amssymb}
\usepackage{amsmath}
\usepackage{dsfont}
\DeclareMathAlphabet{\pazccal}{OMS}{zplm}{m}{n}

\newcommand{\indic}{\mathds{1}} %indicatrice
\DeclareMathOperator{\Cov}{Cov} %covariance
\DeclareMathOperator{\Var}{Var} %variance
\DeclareMathOperator{\Sp}{Sp} %span
\DeclareMathOperator{\Res}{Res} %residue
\DeclareMathOperator{\argmin}{argmin} %argmin
\DeclareMathOperator{\sgn}{sgn} %sign

\usepackage{amsthm}% The amsthm package provides 
\usepackage[hang,flushmargin]{footmisc} 
\usepackage{graphics}
\usepackage{graphicx} % Include figure files
\usepackage{subcaption}
\usepackage{mathrsfs}
\usepackage{stmaryrd}
\usepackage{bbm}
\usepackage{enumitem}
\usepackage{booktabs}
\usepackage{sidecap}
\usepackage{xparse}
\usepackage{multirow}

\usepackage[margin=3cm]{geometry} 

\theoremstyle{plain}

\newtheorem{prop}{Proposition}[section]
\newtheorem{thm}{Theorem}[section]

\setcitestyle{square}

\journal{---}%Finance and Stochastics}
\begin{document}
\begin{frontmatter}
%% Title, authors and addresses

%% use the tnoteref command within \title for footnotes;
%% use the tnotetext command for theassociated footnote;
%% use the fnref command within \author or \affiliation for footnotes;
%% use the fntext command for theassociated footnote;
%% use the corref command within \author for corresponding author footnotes;
%% use the cortext command for theassociated footnote;
%% use the ead command for the email address,
%% and the form \ead[url] for the home page:
%% \title{Title\tnoteref{label1}}
%% \tnotetext[label1]{}
%% \author{Name\corref{cor1}\fnref{label2}}
%% \ead{email address}
%% \ead[url]{home page}
%% \fntext[label2]{}
%% \cortext[cor1]{}
%% \affiliation{organization={},
%%            addressline={}, 
%%            city={},
%%            postcode={}, 
%%            state={},
%%            country={}}
%% \fntext[label3]{}

%\title{Market information in the rough volatility paradigm}
\title{Market information of the fractional stochastic regularity model}
\author[inst1]{Daniele Angelini}

\affiliation[inst1]{organization={MEMOTEF, Sapienza University of Rome},Department and Organization
            %addressline={ }, 
            %city={ },
            %postcode={00100}, 
            %state={State One},
            country={Italy}}

\author[inst2]{Matthieu Garcin}
\affiliation[inst2]{organization={De Vinci Higher Education, De Vinci Research Center, Paris},Department and Organization
            country={France}}

\begin{abstract}
The \textit{Fractional Stochastic Regularity Model} (FSRM) is an extension of Black-Scholes model describing the multifractal nature of prices. It is based on a multifractional process with a random Hurst exponent $H_t$, driven by a \textit{fractional Ornstein-Uhlenbeck} (fOU) process. When the regularity parameter $H_t$ is equal to $1/2$, the efficient market hypothesis holds, but when $H_t\neq 1/2$ past price returns contain some information on a future trend or mean-reversion of the log-price process. In this paper, we investigate some properties of the fOU process and, thanks to information theory and Shannon's entropy, we determine theoretically the serial information of the regularity process $H_t$ of the FSRM, giving some insight into one's ability to forecast future price increments and to build statistical arbitrages with this model. An application to the forecast of future daily price returns of major stock indices shows promising results.
\end{abstract}
\begin{keyword}
%% keywords here, in the form: keyword \sep keyword
Fractional Ornstein-Uhlenbeck process \sep Hurst exponent \sep Shannon entropy \sep serial information \sep nonlinear serial dependence
%% PACS codes here, in the form: \PACS code \sep code
%\PACS 89.65.Gh \sep 02.50.Ey \sep 02.60.Ed
%% MSC codes here, in the form: \MSC code \sep code
%% or \MSC[2008] code \sep code (2000 is the default)
%\MSC 60G17 \sep 60G22 \sep 62J05 \sep 91G15

%JEL classification:
%G14 \sep
%C22 \sep
%C58 
\end{keyword}

\end{frontmatter}

%% \linenumbers
%% main text
\section{Introduction} \label{sec:intro}
In financial mathematics, the most famous model for option pricing is the Black-Scholes model~\cite{BlackScholes1973,Merton1973}, which, under the no-arbitrage assumption, describes the price dynamics $P_t$ of an underlying asset by means of the stochastic differential equation
\begin{equation}\label{eq:BlackScholes}
    \frac{dP_t}{P_t} = r dt + \sigma dW_t,
\end{equation}
where $W_t$ is a standard Brownian motion. Outside the risk-neutral framework, the study of the stylized facts of price returns, among which self-similarity and long-range dependence~\cite{Ding1993,Evertsz1995b,RamseyUsikovZaslavsky1995,Cont2001}, has aroused in finance some interest in fractional processes such as the \textit{fractional Brownian motion} (fBm)~\cite{MandelbrotVanNess1968,Cont2005}. An fBm $B^H_t$, for a Hurst exponent $H\in(0,1/2]$ (respectively $[1/2,1)$), is the fractional derivative (resp. integral) of order $1/2-H$ (resp. $H-1/2$) of a standard Brownian motion. By substituting the Brownian measure $dW_t$ in equation~\eqref{eq:BlackScholes} by the fractional measure $dB_t^H$, we obtain the fractional Black-Scholes model, in which one can adjust the serial dependence of the returns and obtain a process that exhibits long-range dependence when $H>1/2$ as well as self-similarity.

Since the fBm is a non-Markovian process, using it for describing log-prices supposes that one can use past prices to profitably forecast future price returns in average, thus contradicting the \textit{Efficient Market Hyphotesis} (EMH)~\cite{Fama1970}. Does it mean that this model also induces pure arbitrage? This overriding question has been the subject of a large literature~\cite{BenderSottinenValkeila2007}. Though pure arbitrages exist, according to this model, when trading in continuous or even in discrete time~\cite{Rogers1997,Cheridito2003}, arbitrage opportunities disappear when one imposes specific transaction costs or a minimal, and possibly extremely small, interval of time between two consecutive transactions~\cite{Cheridito2003,Guasoni2006}. This last condition reflects the reality of frictions in financial markets, so that one cannot argue from the no-arbitrage condition to discard the fBm for modelling log-prices~\cite{Cont2005}. On the other hand, statistical arbitrages are still possible with this model as soon as $H\neq 1/2$~\cite{GuasoniNikaRasonyi2019,GuasoniMishuraRasonyi2021,Garcin2022}. Depending on the value of the Hurst exponent, one can indeed make predictions of future increments of this process, based on conditional expectations~\cite{NuzmanPoor2000}: when $H>1/2$ successive increments are positively correlated, when $H<1/2$, they are negatively correlated.

In the perspective of accurately evaluating the propensity of a model to induce statistical arbitrage, it seems important to quantify the information contained in past observations of this process with respect to its future evolution. Probabilistic information theory provides useful tools for addressing this question, based on Shannon's entropy~\cite{Shannon1948}. It states that uncertainty and information depend on the shape of a probability distribution: with a uniform distribution we have zero information and with a Dirac distribution we have maximum information~\cite{CoverThomas1991,Garcin2023}. After a binarization of the data using the sign of price returns, like in the Risso's method~\cite{Risso2008}, one can quantify the information contained in past price returns, along with a statistical test of non-zero information~\cite{Shternshis2022,BroutyGarcin2023}. Assuming that log-prices follow an fBm, it is also possible to have a theoretical expression for this information~\cite{BroutyGarcin2024}.

Many possible extensions of the fBm have been investigated for modelling log-prices, in order to depict other empirical properties, such as stationarity~\cite{Zili2017,Garcin2019}, if one considers for example foreign-exchange rates~\cite{GuasoniMishuraRasonyi2021}, fat tails, with fractional stable processes~\cite{SamorodnitskyTaqqu1994,Stoev2002,AmmyDriss2023}, or time-varying Hurst exponents, as in the \textit{multifractional Brownian motion} (mBm)~\cite{Coeurjolly2005,PeltierVehel1995,Garcin2017} or in the \textit{Generalized multifractional Brownian motion} (GmBm)~\cite{Ayache2000,AyacheVehel1999,AyacheVehel2000}. In these last two models, the regularity parameter $H_t$ is a deterministic function of time. Another specification is put forward in the \textit{Multifractional Process with Random Exponent} (MPRE)~\cite{Ayache2013,AyacheBouly2022,AyacheEsserHamonier2018,AyacheTaqqu2005,LobodaMiesSteland2021}, in which $H_t$ is a stochastic process. 

The MPRE has found some applications in finance~\cite{BPP2012,BianchiPianese2018,Garcin2019b}. Indeed, just as stochastic volatility models extend the Black-Scholes model by replacing the constant volatility parameter by a stochastic process, the \textit{Fractional Stochastic Regularity Model} (FSRM) extends the fractional Black-Scholes model by replacing the constant Hurst exponent by a stochastic process~\cite{AngeliniBianchi2023}. In the FSRM, the process $H_t$ is specified as a stationary \textit{fractional Ornstein-Uhlenbeck} (fOU) process, the fractional extension of an Ornstein-Uhlenbeck process~\cite{CheriditoKawaguchiMaegima2003}. The FSRM thus writes as follows:
\begin{equation}\label{eq:FSRP}
    \begin{cases}
        \log(P_t) = B_t^{H_t,C}\\
        H_t = \mathcal H + \eta\int_{-\infty}^te^{-\lambda(t-s)}dB_s^H,
    \end{cases}
\end{equation}
where the log-price is described by an MPRE $B_t^{H_t,C}$ of scale parameter $C>0$ and random Hurst exponent $H_t$, which is itself a fOU of long-term average $\mathcal H\in(0,1)$ and parameters $\eta,\lambda>0$. The regularity of the fOU process, $H_t$, is driven by the fractional measure $dB^H_s$, where $B^H_s$ is an fBm with a constant Hurst exponent $H\in(0, 1)$.

From a statistical arbitrage perspective, the ability to forecast future values of the Hurst exponent $H_t$ offers significant advantages in financial trading. When $H_t > 1/2$, price increments exhibit positive autocorrelation, suggesting that past price trends are likely to continue. This phenomenon supports momentum trading strategies, where traders buy assets that have performed well in the recent past, expecting that the upward trend will persist. Conversely, when $H_t<1/2$, price increments display negative autocorrelation, implying that prices are more likely to reverse direction, which aligns with mean-reverting strategies. In this case, traders would sell assets that have risen in value, anticipating a correction or return to a lower equilibrium price~\cite{BianchiPianese2018}. Given these dynamics, accurately forecasting $H_t$ allows traders to adapt their strategies based on whether the market is in a trending phase (momentum) or a correction phase (mean-reversion). The ability to anticipate changes in $H_t$ could thus provide a competitive edge, allowing traders to capitalize on price movements that deviate from pure randomness, as suggested by traditional models that assume market efficiency. However, forecasting $H_t$ is not trivial. The process itself is driven by complex stochastic dynamics, and its value fluctuates over time~\cite{BianchiPianese2018,AngeliniBianchi2023}. The focus of this paper is to determine whether $H_t$ can indeed be forecast, and if so, how the information embedded in past price movements can be leveraged to predict future values of $H_t$.

Using information theory and assuming that the regularity $H_t$ is modelled by a fOU process, we establish theoretically the serial dependence contained in such a fOU process and summarize it in a quantity called serial information.\footnote{ We call market information the serial information contained in a series of prices increments.} Such an approach has already been followed for another stationary fractional process~\cite{BroutyGarcin2024}, namely the delampertized fBm~\cite{Flandrin2003,CheriditoKawaguchiMaegima2003}, but with a focus on future increments of the process instead of simply future values, as we are interested in here. In the case of the fBm process, a zero information has been observed only for $H=1/2$. Instead, for a delampertized fBm, two different regimes appear: in the fractal regime, that is when the mean-reverting strength tends to zero, a behaviour similar to the fBm is obtained again; in the stationary regime, that is for a stronger mean-reversion, the parameter $H=1/2$ leads to a very high serial information~\cite{BroutyGarcin2024}. In our work, knowing the similarities between a fOU and a delampertized fBm~\cite{CheriditoKawaguchiMaegima2003}, we also expect two different regimes for the information. The main difference between our approach and the existing method applied to the delampertized fBm~\cite{BroutyGarcin2024} is that our fractional stationary process does not directly describes the price. This has a consequence in the way we build binary distributions. Indeed, in the latter work, the information relies on the binarization of the increments of the process, whereas in our article the binarization is applied to the fOU process instead of to its increments. In our financial perspective, the sign of an increment of $H_t$ is thus less important than the sign of $H_t-1/2$, since the latter is directly related to one's ability to forecast future price returns using equation~\eqref{eq:FSRP}.

In order to justify the use of the FSRM and the relevance of the informational study, we derive a forecasting procedure and apply it to three stock indices. We obtain promising results with good performance of the predictor of future price increments when filtering the forecast. More precisely, for some range of parameters, we know \textit{ex ante} that our forecast is uncertain. So, we only focus on forecasts in which we are confident enough. This happens roughly $14\%$ of the days for S\&P 500 index, $100\%$ for Dow Jones index, and $38\%$ for NYSE Composite index, with hit rates close to $60\%$ for these selected days.

The paper is structured as follows. In Section \ref{sec:Reg} we introduce the FSRM and the fOU process, with some of its properties. In Section \ref{sec:marketInf} we explain how one can use information theory, particularly Shannon's entropy, to measure nonlinear serial dependence. In Section \ref{sec:serialdep} we study the serial dependence contained in the regularity process $H_t$ of the FSRM, deriving the serial information of the fOU process as well as the conditional probability of its future value. Section~\ref{sec:estimforecast} presents together the estimation and forecasting procedure. An application to financial data is exposed in Section~\ref{sec:empirical}. Section \ref{sec:conclusion} concludes.

\section{Regularity modelling} \label{sec:Reg}

The FSRM assumes a multifractal behaviour of the price process, with a random regularity parameter following a fOU process. In this section, we derive successively some properties of the FSRM and of the fOU process. Finally, beyond the multifractal feature, we provide another interpretation of the FSRM, related to stochastic volatility models.

\subsection{Fractional stochastic regularity model}\label{sec:2.1}

As a core concept in the FSRM, we first introduce an fBm in the moving-average representation~\cite{MandelbrotVanNess1968,Coeurjolly2001} 
$$B_t^{H,C}=\frac{C\sqrt{\Gamma(2H+1)\sin(\pi H)}}{\Gamma(H+1/2)}\int_{\mathbb{R}}\Big[(t-u)_+^{H-1/2}-(-u)_+^{H-1/2}\Big]dW_u,$$
where $H\in(0,1)$ is the Hurst parameter, $x_+ = \max(0,x)$, $W_t$ is a standard Brownian motion, and $C$ is a scale parameter equal to the variance of an increment of duration 1 of the fBm. When $C=1$, we simply write $B_{t}^H=B_{t}^{H,C}$. This process has stationary and self-similar increments, with
\begin{equation}\label{eq:VarIncr_fbm}
    \mathbb{E}\Big[(B_{t}^{H,C}-B_{u}^{H,C})^2\Big] = C^2\vert t-u\vert^{2H},
\end{equation}
for $t,u\in\mathbb{R}$. The Hurst parameter is related to the H\"older regularity of $B_{t}^{H,C}$: the greater $H$, the smoother $B_{t}^{H,C}$.

In financial applications, a constant Hurst exponent is often too limiting because of a multifractal feature of prices. Therefore we need to introduce another process, whose regularity varies through time. This is the purpose of the FSRM, which uses an MPRE~\cite{Ayache2013,AyacheEsserHamonier2018,AyacheTaqqu2005}, that is a multifractional process in which the Hurst-H\"older exponent $H_t$ is itself a stochastic process. The general form of the MPRE also admits a moving-average representation, with the following It\^o integral~\cite{AyacheBouly2022,LobodaMiesSteland2021}:
$$\int_{-\infty}^t k_u(t)dW_u.$$
In the FSRM, we focus on a specific MPRE, with kernel function $k_u(t) = C\big[(t-u)_+^{H_u-1/2}-(-u)_+^{H_u-1/2}\big]$, which satisfies some conditions regarding its differentiability~\cite{AngeliniBianchi2023}, leading to a natural extension of the fBm:
$$B_t^{H_t,C} = C\int_{-\infty}^t \Big[(t-u)_+^{H_u-1/2}-(-u)_+^{H_u-1/2}\Big]dW_u.$$
Setting $m\in\mathbb{R}$ and $\eta,\lambda>0$, we can define the FSRM as in equation \eqref{eq:FSRP}, where the log-prices are modelled by an MPRE whose time-varying Hurst-H\"older parameter is a fOU with a Hurst exponent $H$. \\
Under the condition that $\sup_{t}H_t<\beta_H([0,1])$, where $\beta_H(J)$ is the uniform H\"older exponent over the non degenerate compact interval $J$, the pointwise H\"older exponent at any time $t^*$ of $B_{t^*}^{H_{t^*},C}$ is almost surely equal to $H_{t^*}$. Then, the MPRE verifies the following locally asymptotic property~\cite{BenassiJaffardRoux1997}:
\begin{equation}\label{eq:Benassi}
    \underset{\epsilon \xrightarrow{}0^+}{\lim}\Bigg(\frac{B_{t+\epsilon u}^{H_{t+\epsilon u},C}-B_t^{H_t,C}}{\epsilon^{H_t}}\Bigg)_{u\in\mathbb{R}} \overset{d}{=} \big(B_u^{H_t,C}\big)_{u\in\mathbb{R}},
\end{equation}
where $\overset{d}{=}$ means equality in distribution. Equation \eqref{eq:Benassi} tells us that, in the neighborhood of any time $t$, $B_t^{H_t,C}$ behaves like an fBm with constant Hurst exponent $H_t$. This has some practical consequences for instance for the estimation of such a process.

In our work, we are particularly interested in the case where the long-term average of $H_t$ is $\mathcal H=1/2$, depicting oscillations of the tangent log-price process around the standard Brownian motion. 

\subsection{Fractional Ornstein-Uhlenbeck process}\label{sec:2.2}

Let $(\Omega,\mathcal F,\mathbb P)$ be a probability space and $\lambda,\eta>0$. We consider the following stochastic differential Langevin-like equation,
\begin{equation}\label{eq:LangevinEq}
    dY_t^H = -\lambda Y_t^Hdt + \eta dB_t^H, \quad t\geq 0,
\end{equation}
driven by an fBm $B_t^H$ of Hurst exponent $H\in(0,1)$. The unique almost surely continuous process that solves equation \eqref{eq:LangevinEq} is the restriction to $t\geq 0$ of the process
$$Y_t^H = \eta e^{-\lambda t}\int_{-\infty}^t e^{\lambda u}dB_u^H, \quad t\in\mathbb R,$$
with the initial condition $Y_0^H = \eta\int_{-\infty}^0 e^{\lambda v}dB_v^H$~\cite{CheriditoKawaguchiMaegima2003}. For any $Y_0^H \in L^0(\Omega)$, the stationary process $(Y_t^H)_{t\geq 0}$ is a fOU with initial condition $Y_0^H$ driven by the Hurst exponent $H$. Contrary to equation~\eqref{eq:FSRP}, we have considered here a long-term average $\mathcal H=0$, in order to simplify the equations, but, obviously, the following results are still valid for $\mathcal H\neq 0$.

We are interested in the autocorrelation function of this process. Surprisingly, though the autocovariance of the fOU process has already been studied for comparison with another kind of stationary process derived from the fBm~\cite{CheriditoKawaguchiMaegima2003}, we have not found any explicit expression of the autocorrelation of the fOU. Of course, from a covariance, one can easily get the definition of a variance and of a correlation. But, in the case of the fOU process, obtaining a concise expression for the correlation requires calculating a particular integral with the residue theorem in the complex plane. In what follows, we thus recall the rationale leading to the expression of the autocovariance of the fOU process and we then provide a new valuable expression for its variance and autocorrelation.

For obtaining the autocovariance of $Y^H_t$, one usually uses the spectral representation of the standard fBm $(B_t^H)_{t\in\mathbb{R}}$, with $0<H<1$\cite{PipirasTaqqu2000,AyacheVehel2000,AyacheTaqqu2005},
$$B_t^H = \dfrac{\sqrt{\Gamma(2H+1)\sin(2H)}}{\sqrt{2\pi}}\int_{\mathbb{R}}\frac{e^{itx}-1}{ix}\vert x\vert^{-(H-1/2)}d\widetilde{B}(x),$$
where $\widetilde{B} = B^{I} + iB^{II}$ is a complex Gaussian measure, such that for any Borel set $A$ of finite Lebesgue measure $\vert A\vert$, we have $B^{I}(A) = B^{I}(-A)$, $B^{II}(A)=-B^{II}(-A)$, and $\mathbb{E}\big[B^{I}(A)\big]^{2}=\mathbb{E}\big[B^{II}(A)\big]^2 = \vert A\vert/2$. Interested in the integration of a function $f$ with respect to a fractional Brownian measure, we introduce the integral linear combination
$$\mathcal{I}^H(f) = \int_{\mathbb{R}} f(u)dB^H_u,$$
where $f(u)$ is a step function defined as $f(u) = \sum_{k=1}^{n} f_k1_{[u_k,u_{k+1})}(u)$, for $u\in\mathbb{R}$ and with $f_k$ and $u_{k+1}>u_k$ real values. The quantity $\mathcal{I}^{H}$ is a Gaussian random variable and, if $\mathcal{D}$ denotes the set of step functions on the real line, then $\{\mathcal{I}^H(f): f\in\mathcal{D}\}$ is a subset of the larger linear space
$$\overline{\Sp}(B^H) = \{X:\mathcal{I}^H(f_n)\xrightarrow{L^2}X, \text{ for some } (f_n)\subset\mathcal{D}\}$$
corresponding to the closure in $L^2(\Omega)$ of the span $\Sp(B^H)$ of the increments of the fBm $B^H$~\cite{PipirasTaqqu2000}. Any element $X\in\overline{\Sp}(B^H)$ is a Gaussian random variable with zero mean and variance
$$\Var(X) = \lim\limits_{n\xrightarrow{}+\infty}\Var(\mathcal{I}^H(f_n)).$$
Therefore we can create a relation between $X$ and an equivalence class of sequences of step functions $(f_n)$ such that $\mathcal{I}^H(f_n)\xrightarrow{}X$ in the $L^2(\Omega)$-sense~\cite{PipirasTaqqu2000}. If $f_X$ is the equivalence class, $X$ is the integral with respect to the fBm on the real line:
$$X = \int_{\mathbb{R}}f_{X}(u)dB^H(u).$$
When $H=1/2$, using the Ito's isometry, it is trivial to observe that the Hilbert space $\overline{\Sp}(B^{1/2})$ and $L^2(\mathbb{R})$ are isometric, i.e. there exists a linear map between these two spaces which preserves inner products~\cite{PipirasTaqqu2000}. In fact for $X,Y\in\overline{\Sp}(B^{1/2})$ there exists unique $f_X,f_Y\in L^2(\mathbb{R})$ such that
$$\mathbb{E}[XY] = \int_{\mathbb{R}}f_X(u)f_Y(u)du.$$
We now suppose we have a set of deterministic functions on the real line $\mathcal{C}$ with an inner product $(f,g)_{\mathcal{C}}=\mathbb{E}\big[\mathcal{I}^H(f)\mathcal{I}^H(g)\big]$ for any $f,g\in\mathcal{D}\subset\mathcal{C}$ and $\mathcal{D}$ dense in $\mathcal{C}$. Then there exists an isometry between $\mathcal{C}$ and a linear subspace of $\overline{\Sp}(B^H)$ \cite[Proposition 2.1(a)]{PipirasTaqqu2000}. An example of such an inner-product space that satisfies all the above conditions has been introduced by Samorodnitsky and Taqqu and is defined by
$$\widetilde{\Lambda}^H = \bigg\{f:f\in L^2(\mathbb{R}), \int_{\mathbb{R}}\vert\widehat{f}(x)\vert^2\vert x\vert^{1-2H}dx\bigg\},$$
where $\widehat{f}$ denotes the Fourier transform of a function $f$, that is $\widehat{f}=\int_{\mathbb{R}}e^{ixu}f(u)du$, with the inner product
$$(f,g)_{\widetilde{\Lambda}^H} = \dfrac{\Gamma(2H+1)\sin(\pi H)}{2\pi}\int_{\mathbb{R}}\widehat{f}(x)\overline{\widehat{g}(x)}\vert x\vert^{1-2H}dx$$
for any functions $f,g$ in the set of step functions $\mathcal{D}$~\cite{SamorodnitskyTaqqu1994}. It has been later noted that, for all $H\in(0,1)$ and $s>0$, the functions $f(x)=1_{\{x\leq0\}}e^{\lambda x}$ and $g(x)=1_{\{x\leq s\}}e^{\lambda x}$ belong to the inner-product space $\widetilde{\Lambda}^H$~\cite{CheriditoKawaguchiMaegima2003}. Therefore, considering that $\widehat{f}(x) = \frac{1}{\lambda-ix}$ and $\overline{\widehat{g}(x)} = \frac{1}{\lambda+ix}e^{(\lambda+ix)s}$, we obtain the covariance function of a fOU~\cite{CheriditoKawaguchiMaegima2003}: $\forall s,t\in\mathbb{R}$,
\begin{equation}\label{eq:cov_fOU}
\begin{array}{ccl}
    \Cov(Y^H_t,Y^H_{t+s}) & = & \eta^2 e^{-\lambda s}(f,g)_{\widetilde{\Lambda}^H} \\
    & = & \eta^2\frac{\Gamma(2H+1)\sin(\pi H)}{2\pi\lambda^{2H}}\int_{-\infty}^{\infty}e^{i\lambda sx}\frac{|x|^{1-2H}}{1+x^2}dx \\
    & = & \eta^2\frac{\Gamma(2H+1)\sin(\pi H)}{\pi\lambda^{2H}}\int_0^{\infty}\cos(\lambda sx)\frac{x^{1-2H}}{1+x^2}dx,
\end{array}
\end{equation}
where the last equality is justified by the fact that the function $x\mapsto |x|^{1-2H}/(1+x^2)$ is even.

Equation~\eqref{eq:cov_fOU} has been used in the literature to show the difference of nature between a fOU process and the Lamperti transform of an fBm~\cite{CheriditoKawaguchiMaegima2003,Garcin2019}. Evaluating this autocovariance in $s=0$ directly provides us with the variance of the process. However, this variance is based on an integral, whose solution is obtained in the following theorem, which, along with the expression of the correlation, will be useful for the rest of the article.

\begin{thm}\label{thm:autocorr_fOU} 
    Let $\lambda,\eta>0$ and $s>0$. The variance and the autocorrelation function of a fOU process $Y_t^H$ are respectively
    $$\Var(Y_t^H) = \frac{\eta^2\Gamma(2H+1)}{2\lambda^{2H}}$$    and
    \begin{equation}\label{eq:autocorr_fOU}
        \rho(Y_t^H,Y_{t+s}^H) = \frac{2\sin(\pi H)}{\pi}\int_0^{\infty}\cos(\lambda sx)\frac{x^{1-2H}}{1+x^2}dx.  
    \end{equation}
\end{thm}

\begin{figure}[!ht]
\centering
    \includegraphics[width=0.8\textwidth]{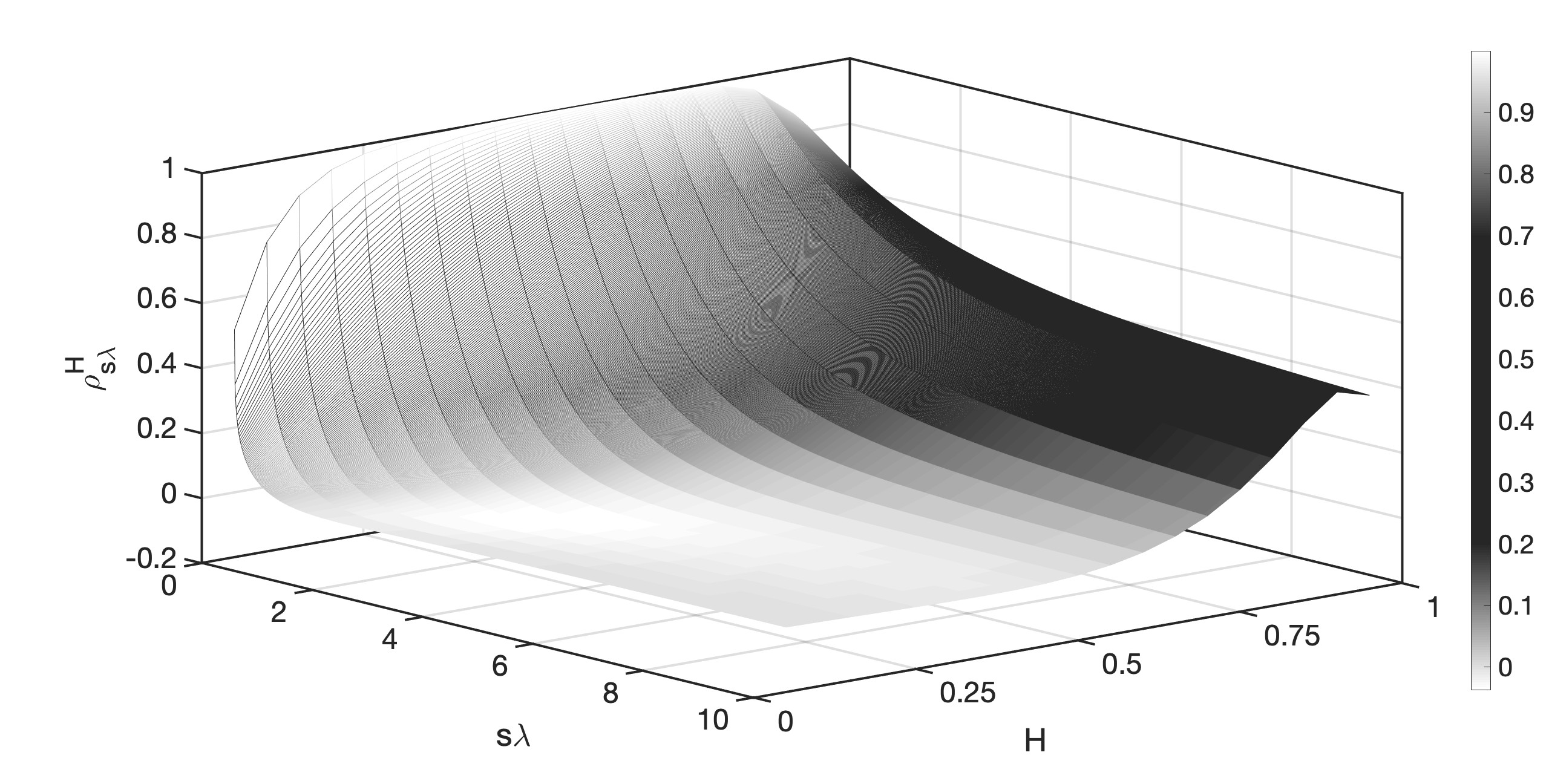}
    \begin{minipage}{0.7\textwidth}\caption{\small Autocorrelation function $\rho_{s\lambda}^H$ for $H\in[0.05,0.95]$ with a step $\Delta H=0.05$, $s\lambda\in[0.01,10]$ with $\Delta s\lambda=0.01$.}
    \label{fig:Autocorr}
    \end{minipage}
\end{figure}

The proof of Theorem~\ref{thm:autocorr_fOU} is postponed in \ref{sec:thm_autocorr_fOU}. We note that the autocorrelation $\rho_{s\lambda}^H=\rho(Y_t^H,Y_{t+s}^H)$ obtained in equation~\eqref{eq:autocorr_fOU} does not distinctly depend on $s$ and $\lambda$ but only on the product $s\lambda$. Figure~\ref{fig:Autocorr} shows this autocorrelation calculated with a trapezoidal integration, as a function of $s\lambda$ and $H$. 

\begin{figure}[!ht]
\centering
    \includegraphics[width=0.48\textwidth]{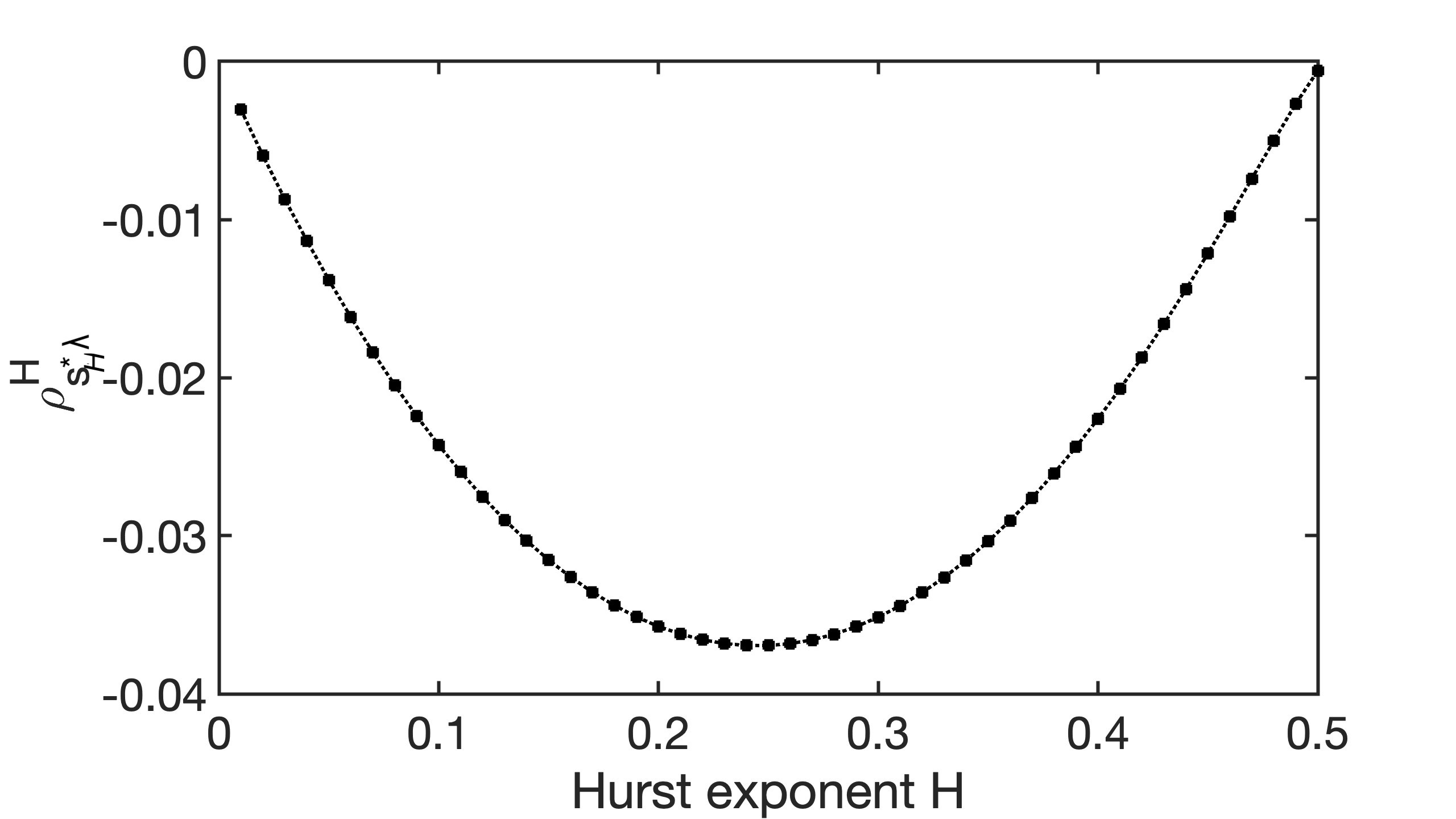}
    \includegraphics[width=0.48\textwidth]{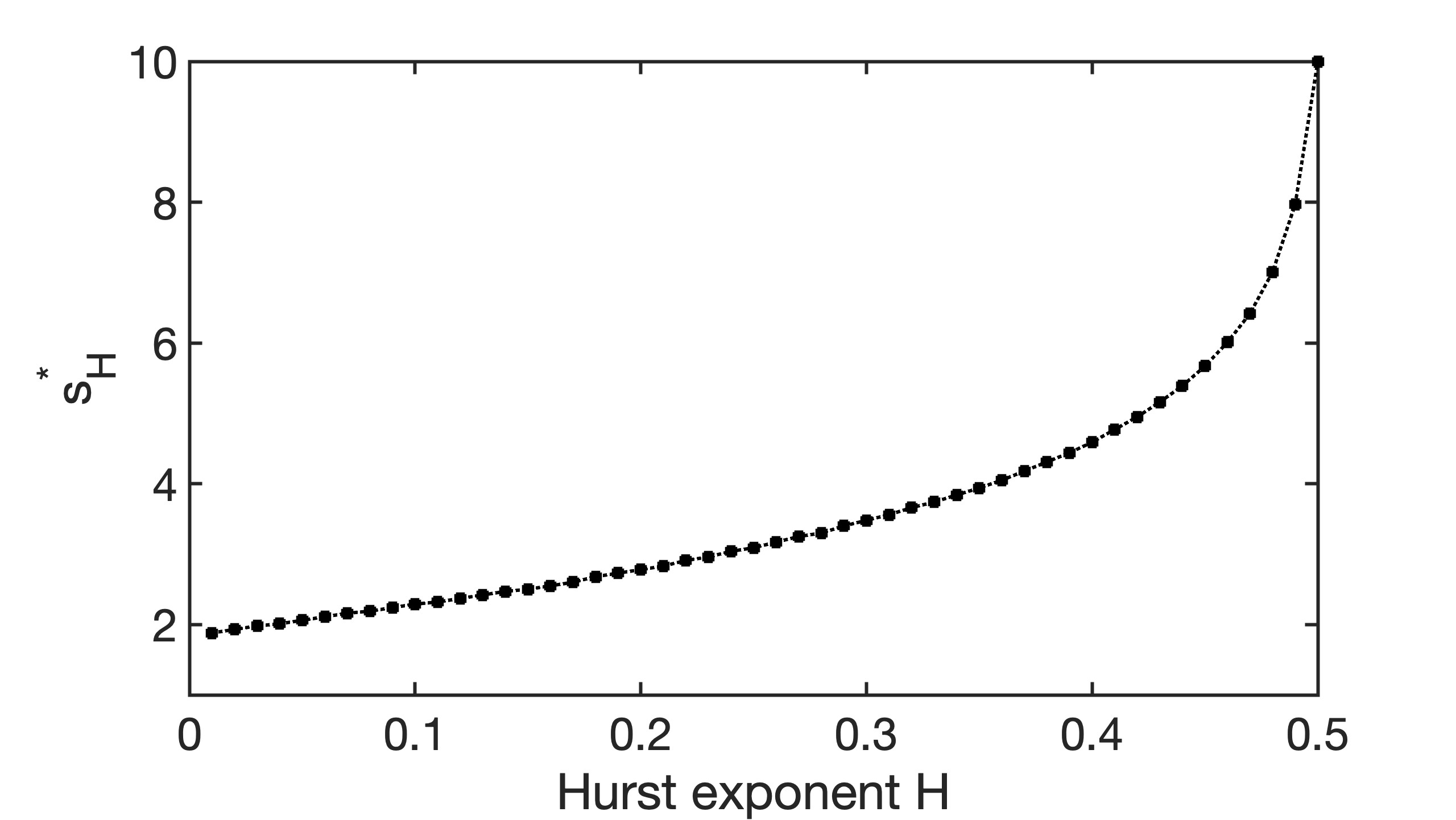}
    \begin{minipage}{0.7\textwidth}\caption{\small  Minimum values of the autocorrelation function $s\lambda\mapsto\rho_{s\lambda}^H$ (left) and corresponding lag $s_H^{\star}$ (right) for $H\in[0.01,0.5]$ with a step $\Delta H=0.01$ and $\lambda=1$.}
    \label{fig:MinAutocorr}
    \end{minipage}
\end{figure}

Setting for example $\lambda=1$ we can study the autocorrelation as a function of the lag $s\in[0.01,10]$. We can see that in the region $H>0.5$ a fOU has a positive autocorrelation, and even a long-range behaviour as prescribed in~\cite{CheriditoKawaguchiMaegima2003}. For $H<0.5$ we observe a short-range positive autocorrelation and, for longer ranges, an anti-persistent behaviour, that is $\rho_{s\lambda}^H<0$. Focusing on this latter case, we display in Figure~\ref{fig:MinAutocorr} the minimum autocorrelation and the corresponding lag 
\begin{equation}\label{eq:s_star}
s_H^{\star}=\argmin_{s\in[0,s_{\text{max}}]} \rho_{s}^H,
\end{equation}
obtained numerically for $s_{\text{max}}=10$. We have a minimum peak of the autocorrelation for $H=0.25$.

\subsection{Financial interpretation}\label{sec:interp}

We have justified the FSRM by the multifractal nature of log-prices. We now present another interpretation of this model that is related to stochastic volatility. Indeed, from equation~\eqref{eq:VarIncr_fbm}, the variance of an increment of duration $s>0$ of an fBm is given by
$$\sigma^2(s) = \mathbb{E}\Big[(B_{t+s}^{H,C}-B_{t}^{H,C})^2\Big] = C^2 s^{2H}.$$
Considering that log-prices are described by the fBm $B_{t}^{H,C}$, then $\sigma(s)$ is the volatility of the log-price increments of duration $s$. Applying a logarithm, we get a linear relation between the Hurst exponent of the log-price and the log-volatility, through time scales. This relation is widely used for estimating the parameters of an fBm:
\begin{equation}\label{eq:linear_fbm}
    \log\sigma(s) = \log C + H\log s.
\end{equation}

Stochastic volatility models are a natural extension of Black-Scholes dynamics introduced in equation~\eqref{eq:BlackScholes} and are justified by empirical observations. The consequence of a stochastic volatility model is that the empirical volatility, for a duration $s$, is time-varying. With this assumption and using equation~\eqref{eq:linear_fbm}, we conclude that the Hurst exponent should be time-varying as well:
\begin{equation}\label{eq:TimeVaryingHurst}
H_t = \frac{\log\sigma_t(s)}{\log s} - \frac{\log C}{\log s}.
\end{equation}
Using equation~\eqref{eq:Benassi}, the MPRE seems to be a good way to define log-prices with time-varying Hurst exponent, with a linear mapping between $H_t$ and $\sigma_t$ as advocated by equation~\eqref{eq:TimeVaryingHurst}. Standard stochastic volatility models should therefore be reasonable choices for our stochastic Hurst exponent.

The most famous stochastic volatility models are certainly the Heston model~\cite{Heston1993} and the SABR model~\cite{HaganKumarLesniewskiWoodward2002}. In order to consider the long memory in the volatility process, Comte and Renault introduced the \textit{Fractional Stochastic Volatility Model} (FSVM)~\cite{ComteRenault1998}, where the log-volatility is modelled by a fOU. Modifying the FSVM, the stationary \textit{Rough Fractional Stochastic Volatility Model} (RFSVM) depicts the volatility as a fOU with a parameter $H\sim 0.1$ intended to catch the rough nature of $\ln\sigma_t$~\cite{AlosLeonVives,GatheralJaissonRosenbaum2018}. Despite some debate about the statistical relevance of the model~\cite{ContDas2024,GarcinGrasselli2022}, there has been an enormous literature on rough volatility, which is now a widespread model~\cite{BayerFrizGatheral2016,FordeZhang2017,Fukasawa2021,FukasawaHorvathTankov2021,Fukasawa2019}. These developments about rough volatility, verified by empirical observations ~\cite{FujiwaraFujisaka2013}, encourage us to consider the fOU as a dynamic describing the time-varying Hurst exponent. This is the purpose of FSRM.

Regarding the robustness of the inference of the global Hurst exponent\footnote{ The global Hurst exponent describes the Hurst exponent of the Hurst exponent process of the log-prices. In other words, the global Hurst exponent is the parameter $H$ appearing in equation~\eqref{eq:FSRP}.} of such a model, it has been shown that many estimators~\cite{TaqquTeverovskyWillinger1995}, relying only on single moments of the distribution of returns, introduce non-linear biases by creating artificial roughness when $H>1/2$~\cite{AngeliniBianchi2023}. However, a recent method based on the Lamperti transform showed that even using the entire distribution of returns, the estimate of the global Hurst exponent is very rough~\cite{BianchiAngeliniPianeseFrezza2023}.

Besides its link with log-volatility, the Hurst exponent of log-prices also has an interesting interpretation since a Hurst exponent of $1/2$ is related to the efficient market hypothesis, whereas values greater or lower than $1/2$ underline the opportunity of statistical arbitrages. Obviously, in the FSRM, the regularity $H_t$, being modelled by a fOU augmented by a long-term average $\mathcal H=1/2$, is not constrained to be in the interval $(0,1)$ as it should be. Therefore, we can work with a new variable $\widetilde{H}_t$ such as
\begin{equation}\label{eq:norm_Ht}
    \widetilde{H}_t = \frac{1}{2} + \frac{1}{\pi}\arctan\Big(H_t-\frac{1}{2}\Big).
\end{equation}
With this transformation the new regularity $\widetilde{H}_t$ is well defined in the interval $(0,1)$.

\section{Information theory for serial dependence} \label{sec:marketInf}

Let $\textbf{X}^L_1 = (X_1,\ldots, X_L)'$ be a multivariate discrete random variable which we can see as a string with $L$ characters. Each character of $\textbf{X}^L_1$ can take a binary value $s \in \{0,1\}$. Therefore, the vector $\textbf{X}^L_1$ has $2^L$ possible configurations, denoted as $s_i^L\in\{0,1\}^{2^L}$, with $i\in\llbracket 1,2^L\rrbracket$. The Shannon's entropy of the vector $\textbf{X}^L_1$, which is the measure of its uncertainty, is defined as
\begin{equation}\label{eq:ShannonEntropy}
    E(\textbf{X}^L_1) = -\sum\limits_{i=1}^{2^L} p^L(s^L_i)\log_2\left(p^L(s^L_i)\right),
\end{equation}
where $p^L(s^L_i)=\mathbb{P}(\textbf{X}^L_1=s_i^L)$~\cite{CoverThomas1991}. The more ordered (respectively disordered) the distribution of $\textbf{X}^L_1$, e.g. a Dirac (resp. uniform) distribution, the smaller (resp. larger) the entropy.

If one considers a binary stationary time series $X_1,...,X_n$, with $n>L$, one can use equation~\eqref{eq:ShannonEntropy} as the starting point to build an indicator of nonlinear serial dependence. The vector $\textbf{X}^L_i$ is a vector of $L$ consecutive observations of the time series, starting in time $i$. We are able to capture the serial dependence of the time series $X_i$ by considering the conditional probabilities $p^L(s^1_j|s^L_i)=\mathbb{P}(X_{.+L}=s^1_j\vert \textbf{X}_{.}^{L}=s_i^L)$, where $j\in\{1,2\}$. One can then write the entropy of an augmented vector of size $L+1$,
\begin{equation}\label{eq:entropy}
E^{L+1} = E(\textbf{X}^{L+1}_.) = -\sum\limits_{i=1}^{2^L}\sum\limits_{j=1}^2 p^L(s^1_j\vert s^L_i)p^L(s_i^L)\log_2\Big(p^L(s^1_j\vert s^L_i)p^L(s_i^L)\Big),
\end{equation}
as well as the conditional entropy,
\begin{equation}\label{eq:conditional_entropy}
    E(X_{.+L}\vert \textbf{X}^L_{.}) = -\sum\limits_{i=1}^{2^L}p^L(s_i^L)\sum\limits_{j=1}^2 p^L(s_j^1\vert s_i^L)\log_2\Big(p^L(s_j^1\vert s_i^L)\Big).
\end{equation}
Using the chain rule, we can decompose the entropy as~\cite[Th.2.2.1]{CoverThomas1991}:
$$E^{L+1} = E^L + E(X_{.+L}\vert \textbf{X}^L_{.}).$$

We can characterize the case without serial dependence in the time series by $p^L(1\vert s_i^L)=p^L(0\vert s_i^L)=1/2$, whatever $i\in\llbracket 1,2^L\rrbracket$. As a consequence, after equation~\eqref{eq:conditional_entropy}, the absence of serial dependence leads to a unit conditional entropy and to the following equation~\cite{BroutyGarcin2023,BroutyGarcin2024}:
\begin{equation}\label{eq:EMH-entropy}
    E^{L+1,\star} = E^{L} + 1,
\end{equation}
where $E^{L+1,\star}$ follows the same definition as $E^{L+1}$ in equation~\eqref{eq:entropy}, but replacing both $p^L(1\vert s_i^L)$ and $p^L(0\vert s_i^L)$ by $1/2$. Finally, the serial information is the difference between the entropy in equation \eqref{eq:EMH-entropy}, which assumes no serial dependence, and the true entropy $E^{L+1}$:
\begin{equation}\label{eq:IM}
    I^{L+1} = E^{L+1,\star} - E^{L+1} = 1 - E(X_{.+L}\vert \textbf{X}^L_{.}).
\end{equation}
Equation \eqref{eq:IM} is always non-negative and is equal to zero if and only if we have maximum uncertainty in the time series, that is serial independence.

\section{Serial dependence of the regularity process in the FSRM}\label{sec:serialdep}

In Section~\ref{sec:marketInf}, we have introduced the notion of serial information, which is equal to zero in the absence of serial dependence. This tool and other extensions of Shannon entropy~\cite{Alvarez2012} have proven useful in finance to reveal statistical arbitrages~\cite{BroutyGarcin2023,BroutyGarcin2024}. When applied to time series of price increments, this serial information is also called market information, since it makes it possible to build statistical tests of market efficiency~\cite{BroutyGarcin2023}. In the FSRM studied in the present article, we don't define the market information in the same way, because of the complexity of the model which makes the theoretical calculation of its serial information intractable. Instead, we claim that the knowledge of future Hurst exponents of the log-price process helps to define statistical arbitrages. Indeed, if the future Hurst exponent is higher (respectively lower) than $1/2$, a trend-following (resp. mean-reverting) strategy should be profitable in average~\cite{Garcin2017,Garcin2022}. In other words, because of our MPRE framework, we base the market information of the FSRM on a serial information of the fOU process. This is the method described below.

\subsection{Serial information of the fOU process}\label{sec:fOU}

We apply the above framework of serial information to the binary time series of regularity indicators defined as follows, for $m>0$ corresponding to the time scale at which the process is to be considered,
\begin{equation}\label{eq:regindic}
J_{m,i} = 
    \begin{cases}
        1, \quad \text{if} \; \widetilde{H}_{mi} -\frac{1}{2} > 0 \\
        0, \quad \text{otherwise,}
    \end{cases}
\end{equation}
where $\widetilde H_t$ follows a fOU process transformed as in equation \eqref{eq:norm_Ht}.

In the next theorem we provide the theoretical expression of the serial information when $L=1$, in a way similar to existing results regarding the serial information of binarized versions of an fBm or of a delampertized fBm~\cite{BroutyGarcin2024}. In this framework with a new time scale $m$, the information we are interested in writes $I_m^{2} = 1 - E(J_{m,.+1}\vert J_{m,.})$, consistently with equation~\eqref{eq:IM}. 

\begin{thm}\label{thm:MI_FOU}
    Let $H_t$ be a fOU of Hurst exponent $H$, long-term average $\mathcal H=1/2$, and $\eta,\lambda>0$ be respectively the diffusion and mean-reverting parameters. Considering the transformation $\widetilde H_t = \frac{1}{2}+\frac{1}{\pi}\arctan\big(H_t-\frac{1}{2}\big)\in(0,1)$ for all $t\in\mathbb R$ and the temporal lag $m>0$, the serial information $I^2_m$, introduced in equation~\eqref{eq:IM} and applied to the series of indicators $J_{m,i}$, introduced in equation~\eqref{eq:regindic}, is equal to
    $$I^2_m = 1 + f\Bigg(\frac{1}{2}-\frac{1}{\pi}\arctan\Bigg(\frac{\rho}{\sqrt{1-\rho^2}}\Bigg)\Bigg) + f\Bigg(\frac{1}{2}+\frac{1}{\pi}\arctan\Bigg(\frac{\rho}{\sqrt{1-\rho^2}}\Bigg)\Bigg),$$
    where $f:x>0\mapsto x\log_2(x)$ and $\rho = \frac{2\sin(\pi H)}{\pi}\int_0^\infty \cos(\lambda mx)\frac{x^{1-2H}}{1+x^2}dx$.
\end{thm}

\begin{figure}[!ht]
\centering
\label{fig:LowFrequencies}
    \includegraphics[width=0.48\textwidth]{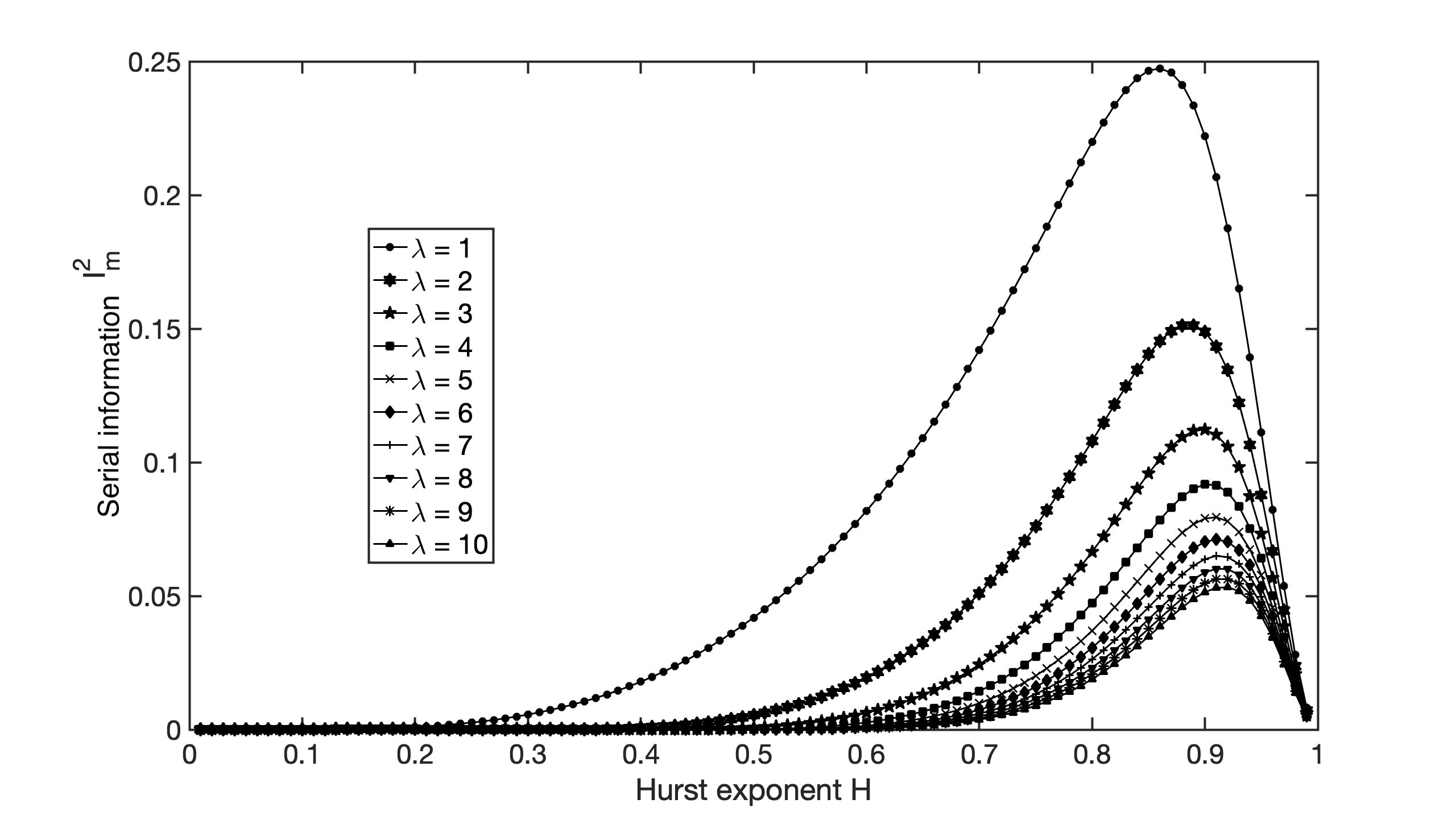}
    \includegraphics[width=0.48\textwidth]{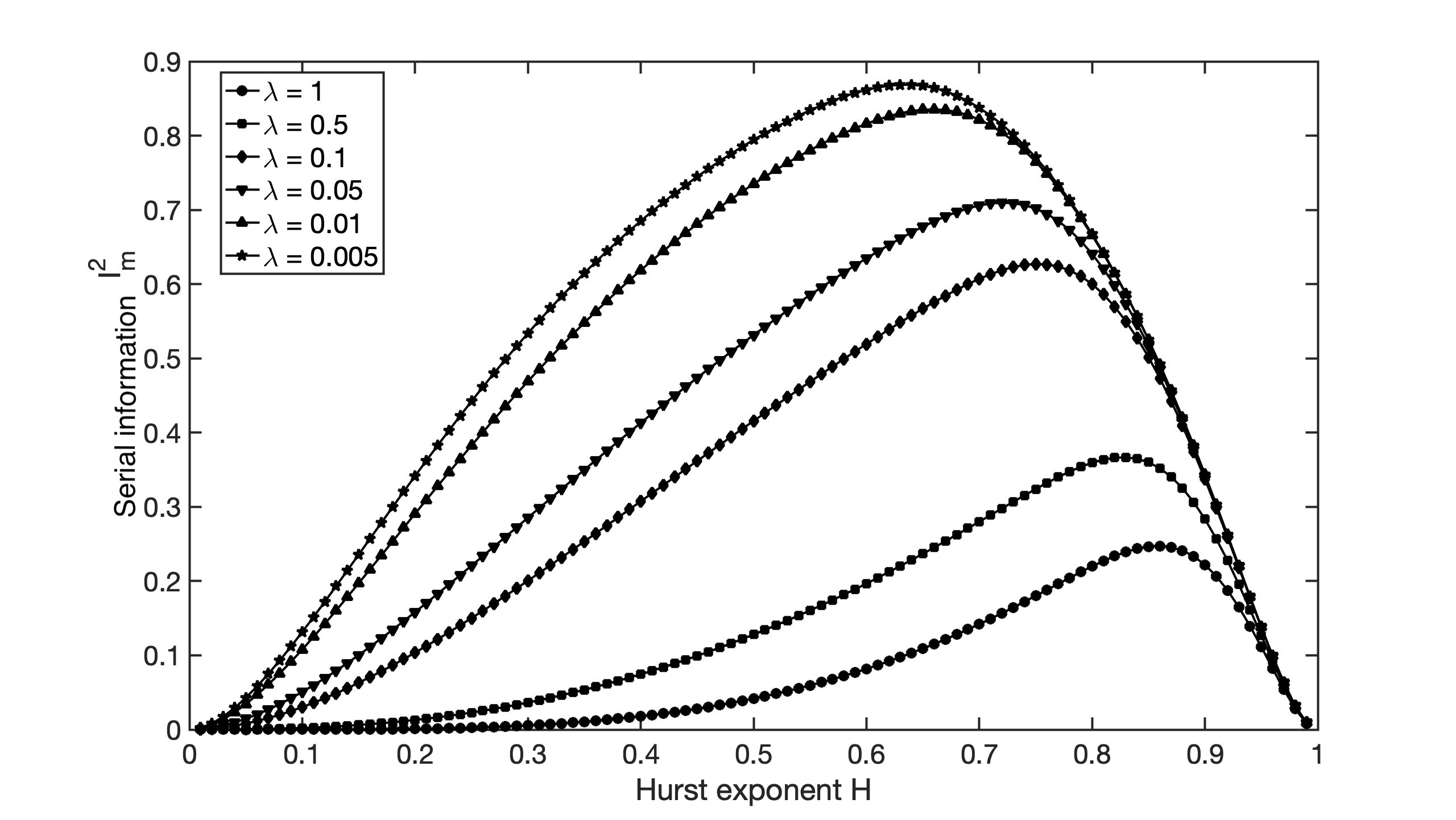} \\
    \includegraphics[width=0.48\textwidth]{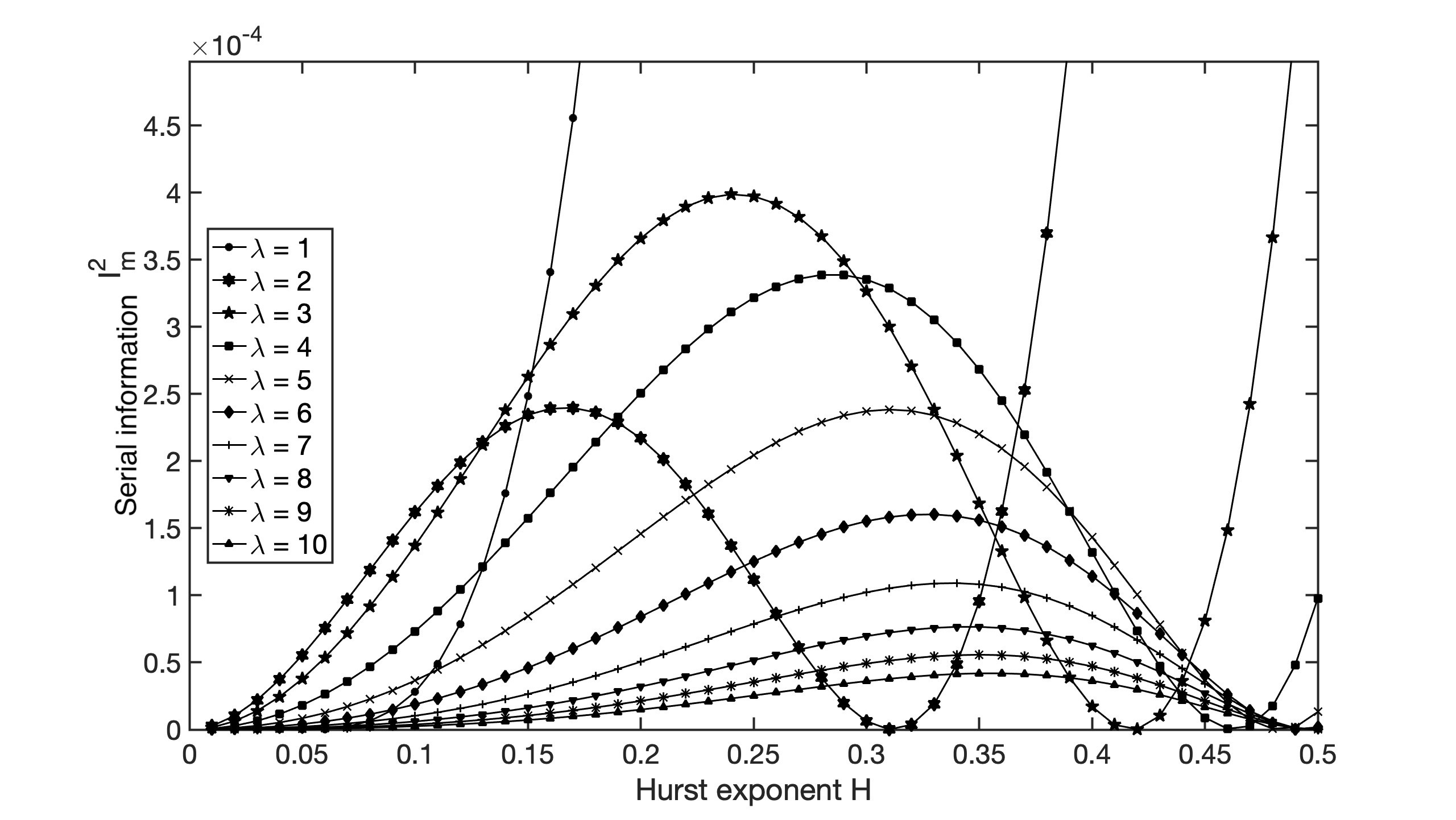}
    \includegraphics[width=0.48\textwidth]{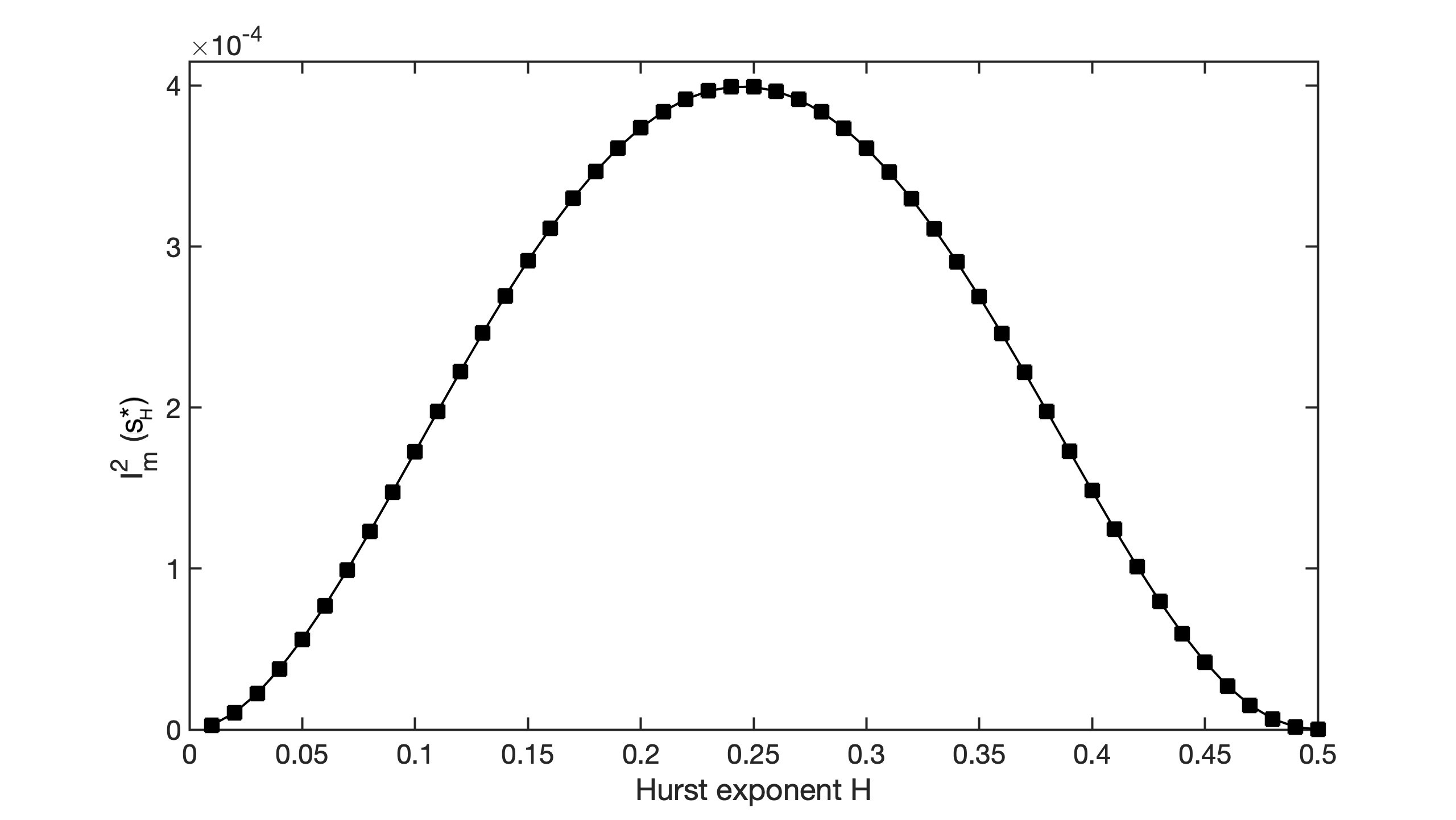}
    \begin{minipage}{0.7\textwidth}\caption{\small Theoretical serial information $I_m^2$ for the binarized fOU process, as a function of the Hurst exponent $H$ of this process, with $m=1$, for the stationary regime ($\lambda>1$, top left and bottom left for a zoom when $H<0.5$) and the fractal regime ($\lambda<1$, top right). The bottom right graph is the theoretical serial information for $\lambda=1$ and a time lag $m=s^{\star}_H$ minimizing the autocorrelation or locally maximizing the information.}
    \label{fig:info_fOU}
    \end{minipage}
\end{figure}

The proof of Theorem~\ref{thm:MI_FOU} is postponed in \ref{sec:thm_MI_FOU}. Since $\rho$ depends on $\lambda m$ and $H$, the information $I_m^2$ depends as well on $\lambda m$ and $H$. For $L=1$, because of the Gaussian nature of the dynamic, the serial information of the fOU process is a simple transformation of the autocorrelation function. It depends on the Hurst parameter $H$ of the fOU process and on the product $m\lambda$. We display this theoretical serial information in Figure~\ref{fig:info_fOU}. We can distinguish two regimes, consistently with the literature on stationary fractional processes~\cite{Garcin2019,Garcin_CNSNS}: a stationary regime when $m\lambda > 1$, that is when considering low-frequency data or strong mean reversion; a fractal regime when $m\lambda < 1$, that is when considering high-frequency data or weak mean reversion.

As one can see in Figure~\ref{fig:info_fOU}, in the stationary regime, the serial information $I^2_m$ admits a peak for a high value of $H$. The curve also moves down when $\lambda$ is higher, indicating a lower serial information of the fOU for stronger mean reversion. In the fractal regime, the lower $\lambda$, the more the peak moves towards the value of $H=1/2$. Moreover, in this case, compared to the stationary regime, the serial information grows when the fOU is very rough, that is when $H\ll 1/2$. It is worth noting that, while a low value of $\lambda$ makes the fOU close to an fBm, we don't get a serial information close to 0 when $H=1/2$. This is because of the way the process in binarized in equation~\eqref{eq:regindic}: an fBm with $H=1/2$ has zero information regarding the sign of a future increment, but it contains some information on its future level compared to a reference value (here $1/2$), as soon as the current level of the process is different from this reference value. This will be illustrated in Section~\ref{sec:probability}.

Because of the similitude between the FSRM and the rough volatility models evoked in Section~\ref{sec:interp}, we are particularly interested in the case $H\ll 1/2$ and we display a zoom on low values of $H$ in the stationary regime in Figure~\ref{fig:info_fOU}. We can observe different peaks, depending on the value $m\lambda$: the maximum peak corresponds to $H=0.25$. These local maxima of the serial information correspond to the minimum autocorrelation of the fOU process. For a fixed mean-reversion strength, we also display this local maximum information obtained for a time lag $m=s^{\star}_H$, with $s^{\star}_H$ defined as in equation~\eqref{eq:s_star}. In short, we can say that in the rough volatility paradigm, when dealing with low-frequency data or very mean-reverting $H_t$ regularities, the market information is very low. The informational content is concentrated in high-frequency data. 

\subsection{Conditional probability of the future regularity}\label{sec:probability}

With regard to price forecasting with the FSRM, we are interested in the probability of obtaining at a future date a regularity greater (or less) than $1/2$, starting from the current Hurst exponent $H_{mi}$. Indeed, in the case where $H_{m(i+1)}>1/2$ (respectively $<1/2$), we will most likely have a future price that follows its past trend (resp. a trend that reverts). Still focusing on the regularity indicator $J_{m,i}$ introduced in equation~\eqref{eq:regindic}, we want to determine the following conditional probability:
\begin{equation}\label{eq:conditional_prob}
    p(1\vert x)=\mathbb{P}(J_{m,i+1} = 1\vert H_{mi}=x) = \mathbb{P}(H_{m(i+1)}>1/2\vert H_{mi} = x).
\end{equation}
Compared to the serial information developed in Section~\ref{sec:fOU}, the conditional probability we now consider in equation~\eqref{eq:conditional_prob} is more granular since the conditioning is not based on the binarized process $J_{m,i}$ but directly on the fOU process.

\begin{prop}\label{prop:1}
Let $H_t$ be a fOU of Hurst exponent $H$, long-term average $\mathcal H=1/2$, and $\eta,\lambda>0$ be respectively the diffusion and mean-reverting parameters. Then, the conditional probability $p(1\vert x)$ introduced in equation~\eqref{eq:conditional_prob} verifies
$$p(1\vert x)=N\left(\frac{\lambda^H\sqrt{2}(x-1/2)}{\eta\sqrt{\Gamma(2H+1)((\rho^H_{m\lambda})^{-2}-1)}}\right),$$
where $N$ is the standard Gaussian cdf and $\rho^H_{m\lambda}=\frac{2\sin(\pi H)}{\pi}\int_0^{\infty}\cos(\lambda mx)\frac{x^{1-2H}}{1+x^2}dx$ is the autocorrelation of a fOU process as introduced in Section~\ref{sec:2.2}.
\end{prop}

The proof of Proposition~\ref{prop:1} is postponed in \ref{sec:prop_1}. We can also easily write this conditional probability for the transformation $\widetilde H_t = \frac{1}{2}+\frac{1}{\pi}\arctan\big(H_t-\frac{1}{2}\big)\in(0,1)$ introduced in equation~\eqref{eq:norm_Ht}: 
$$\mathbb P(J_{m,i+1}|\widetilde H_{mi}=x)=p\left(1\left|\frac{1}{2}+\tan\left(\pi\left(x-\frac{1}{2}\right)\right)\right.\right).$$ 

\begin{figure}[!ht]
    \centering 
    \begin{subfigure}[b]{0.46\textwidth}
    \centering
    \includegraphics[width=1\linewidth]{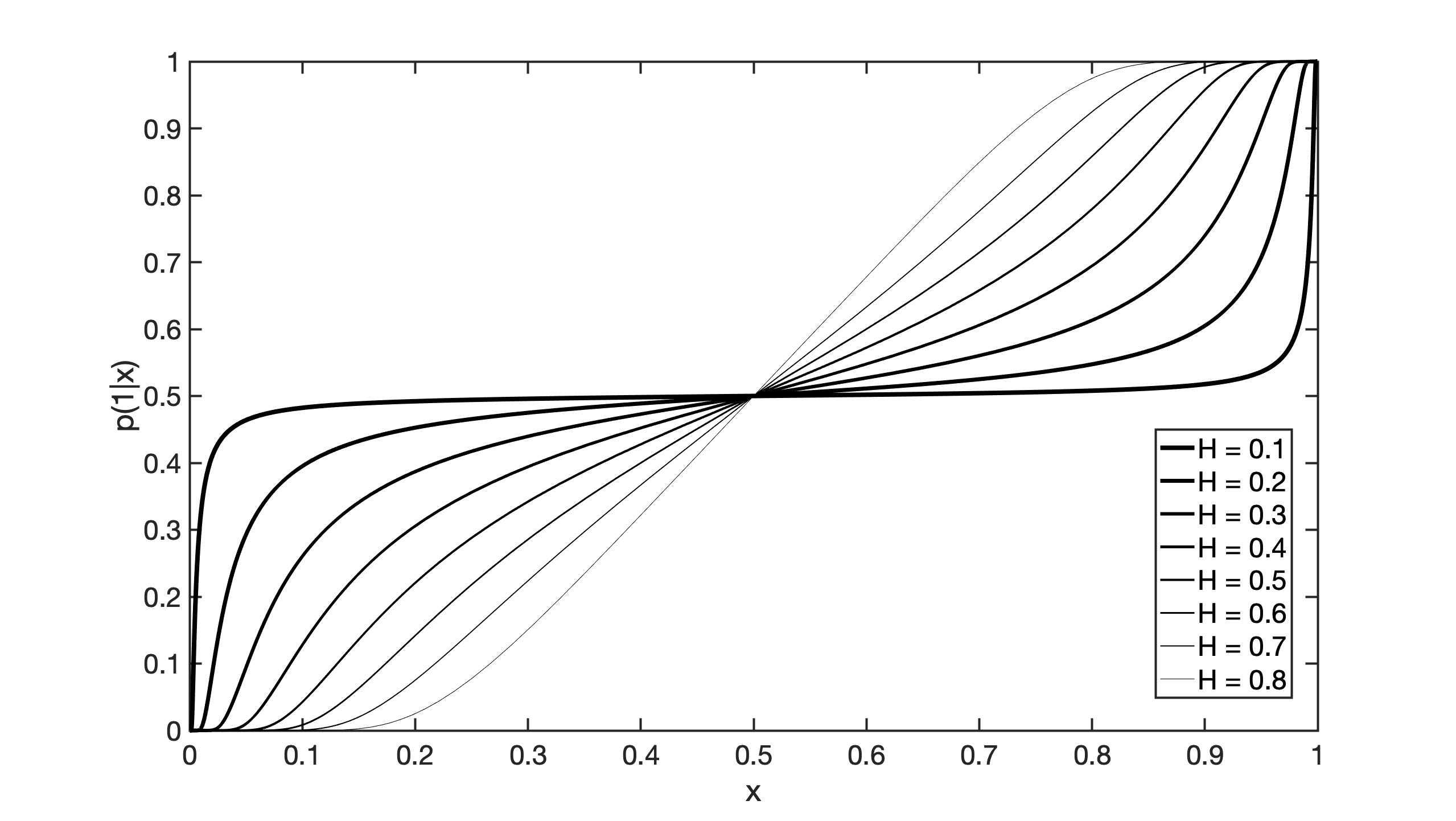}
    \caption{\small }
    \label{fig:prob_H}
    \end{subfigure}
    \hspace{0.05\textwidth}
    \begin{subfigure}[b]{0.46\textwidth}
    \centering
    \includegraphics[width=1\linewidth]{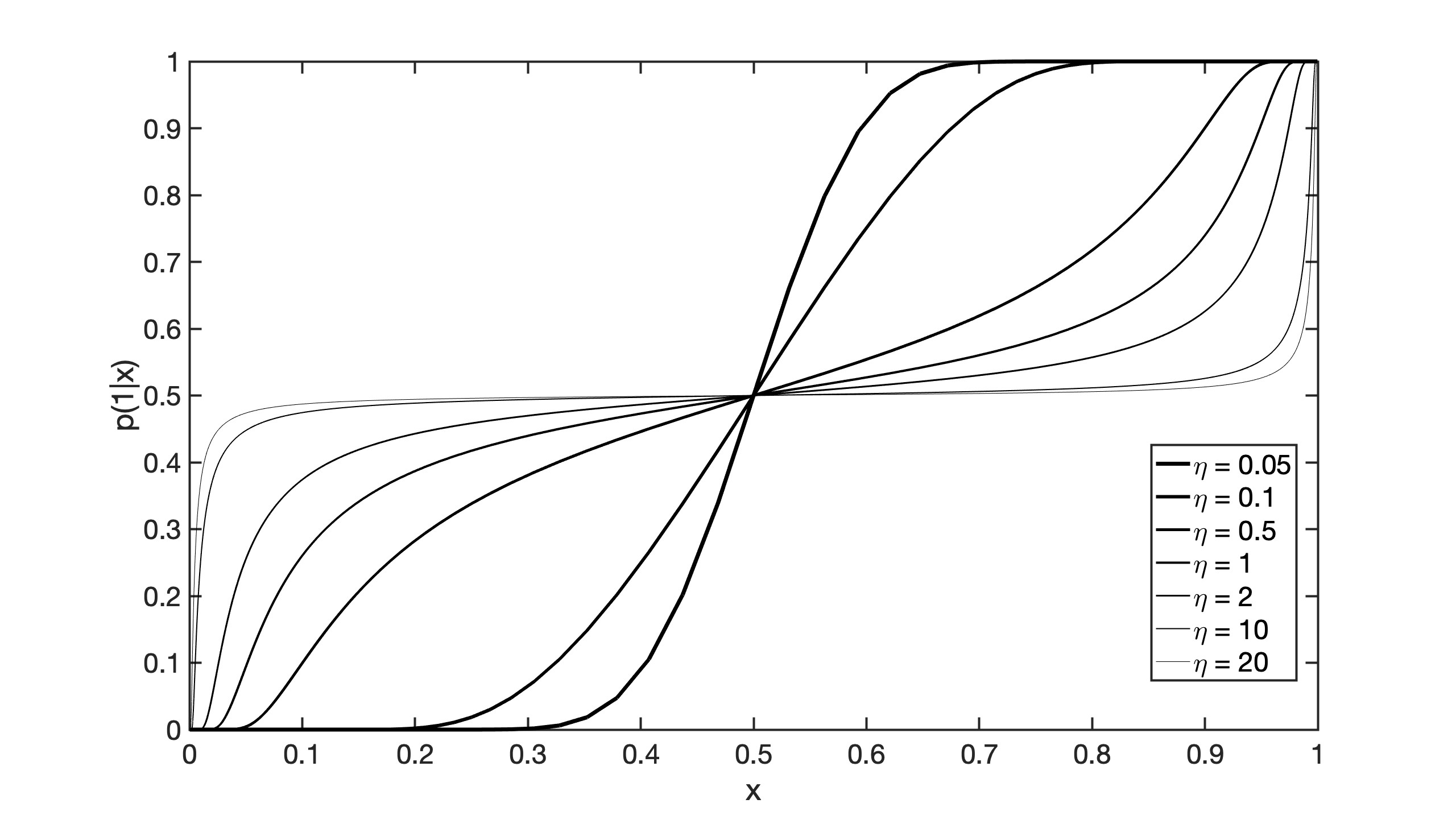}
    \caption{}
    \label{fig:prob_E}
    \end{subfigure}  
\\
    \centering
    \begin{subfigure}[b]{0.46\textwidth}
    \centering
    \includegraphics[width=1\linewidth]{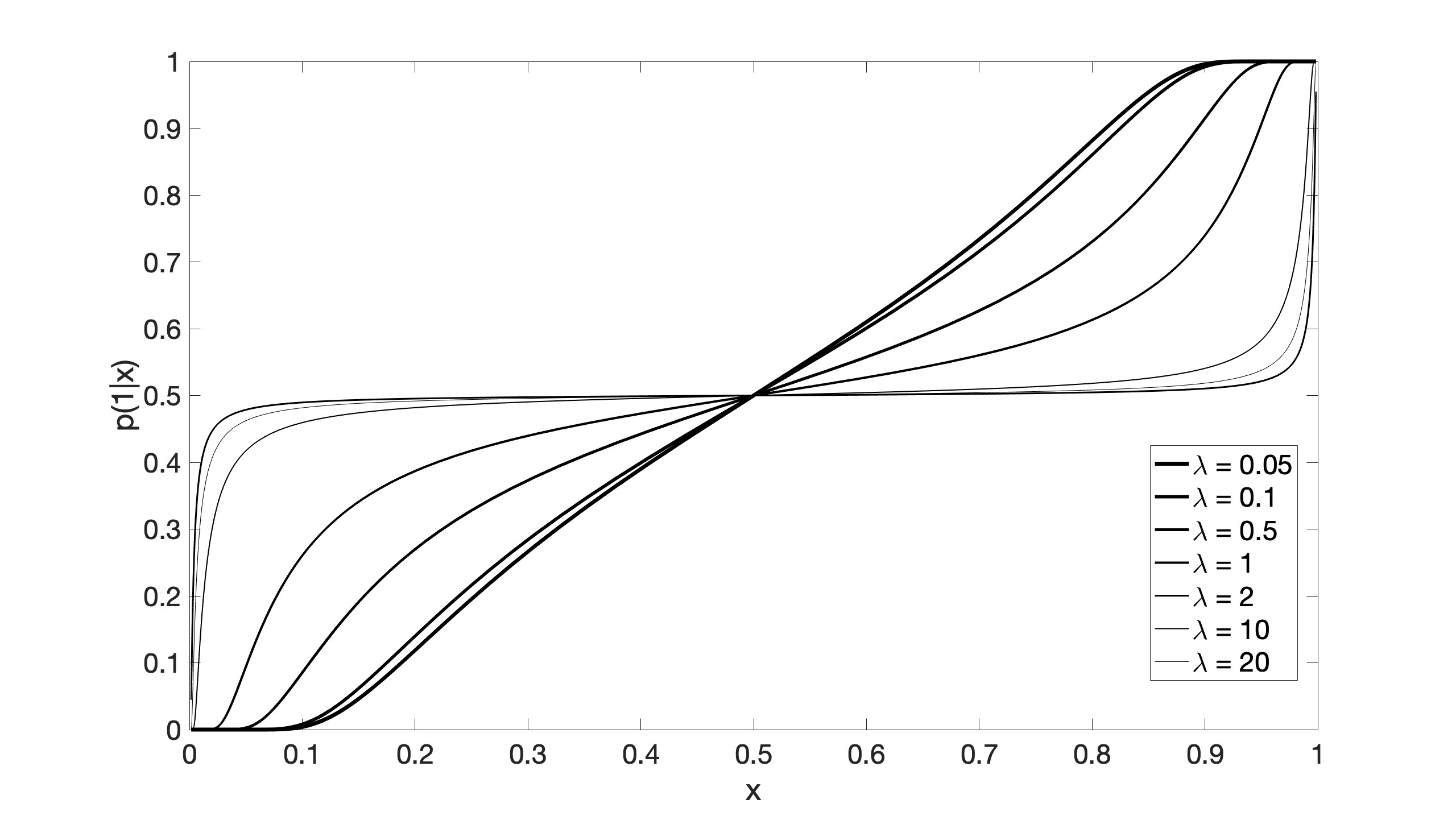}
    \caption{\small }
    \label{fig:prob_L}
    \end{subfigure}
    \caption{ Probability $p(1|x)=\mathbb{P}(H_{m(i+1)}>1/2\vert H_{mi} = x)$ versus the current level of the process $H_{mi} = x$, for various sets of parameters of the fOU: (a) varying the Hurst exponent $H$, setting $\eta = 1$ and $\lambda = 1$; (b) varying the diffusion parameter $\eta$, setting $H = 0.5$ and $\lambda = 1$; (c) varying the mean-reversion parameter $\lambda$, setting $H=0.5$ and $\eta = 1$.}
    \label{fig:Prob1}
\end{figure}

Using Proposition \ref{prop:1}, we display in Figure \ref{fig:Prob1} the probability $p(1|x)$, setting $m=1$, versus the current value $H_{mi}=x$ of the regularity process modelled by a fOU. In panel \ref{fig:prob_H}, setting $\eta=1$ and $\lambda=1$, we observe that the smaller the global $H$ of the fOU, the greater the uncertainty about the value of $H_{i+m}$: in particular, for $H=0.1$, for almost all values of $H_{mi}$, we have $p(1\vert x)\approx 1/2$, whereas, for $H=0.5$, we are far from uncertainty, say with $p(1|x)\leq 0.4$ or $\geq 0.6$, as soon as $H_{mi}\notin[0.35,0.65]$. In panel \ref{fig:prob_E}, setting $H=0.3$ and $\lambda=1$, we plot $p(1\vert x)$ for various values of the diffusion parameter $\eta$: the larger the value of $\eta$, the greater the uncertainty about the future value of the regularity. Finally, in panel \ref{fig:prob_L}, setting $H=0.3$ and $\eta = 1$, we study $p(1\vert x)$ for various intensities $\lambda$ of the mean reversion. Higher values of $\lambda$ in general lead to higher uncertainty, that is to $p(1\vert x)$ closer to $1/2$ for a large range of $x$. To summarize, the quality of the forecast in the FSRM is improved when $H$ is large, when $\eta$ and $\lambda$ are small, and when the current regularity $H_{mi}$ is far from $1/2$.

\section{Forecasting prices with FSRM: a practical approach}\label{sec:estimforecast}

We have explored the informational properties of the fOU process in Section~\ref{sec:serialdep}. This can be used to build trading strategies in the FSRM model. Indeed, in this MPRE setting, the Hurst exponent $H_t$ of the log-price follows a fOU, so that being able to forecast $H_{t+\tau}$ will help us to determine if the log-price process between $t$ and $t+\tau$ will be more mean-reverting or more trend-following.

In what follows, we present successively the estimation procedure, the forecasting procedure, and the method for evaluating the quality of the forecast.

\subsection{Estimation procedure}\label{sec:estim}

The time series of regularity $H_t$ is never directly observable and the only observable data in our possession are prices. Therefore, if the fundamental building block of our approach is the probability $p(1\vert x)$, we need to estimate the parameters $H$, $\lambda$, and $\eta$ of the fOU process that drives the regularity, by applying some estimators to the price series. We will operate with a double estimation process. First, we apply a moving window estimator to the log-prices to obtain the time-varying regularity $H_t$. Then, we use estimators for the parameters $H$, $\lambda$, and $\eta$ to the estimated $H_t$, which is supposed to follow a fOU.

In detail, among the many methods available to estimate a time-varying Hurst exponent, methods based on the variance of first-order or second-order increments are widespread~\cite{IstasLang1997,KentWood1997}. Basically, for a time series $X_1,...,X_N$, it is common to use a sliding window of length $\nu$ and infer a local Hurst exponent in each window by considering the logarithm of a ratio of variances at two time scales, finally obtaining a time series of Hurst exponents: 
\begin{equation}\label{eq:estimH}
\widehat H_t^{\nu}=\frac{1}{2}\log_2\left(\frac{M'_{\nu}(t;X)}{M_{\nu}(t;X)}\right),
\end{equation}
with
$$M_{\nu}(t;X)=\frac{1}{\nu}\sum_{i=0}^{\nu-1} \left|X_{t-i}-2X_{t-1-i}+X_{t-2-i}\right|^2$$
and
$$M'_{\nu}(t;X)=\frac{2}{\nu}\sum_{i=0}^{(\nu-1)/2} \left|X_{t-2i}-2X_{t-2(1+i)}+X_{t-2(2+i)}\right|^2.$$
The obtained time series of estimators $\widehat H_t^{\nu}$ is polluted by measurement noise and one can think about trying to smooth it to diminish the amplitude of the noise~\cite{grech2008,MoralesDiMatteoGramaticaAste2012,Garcin2017}. However, in our framework, the time series of Hurst exponents is not supposed to be smooth because we do consider that log-prices are an MPRE instead of an mBm. Therefore, we don't smooth the time series, and we can instead either use large time windows, using for example intraday data, or use biased estimators with a faster rate of convergence and a technique to estimate and remove the bias~\cite{PianeseBianchiPalazzo2018,AngeliniBianchi2023}.

Once the series $\widehat H_t^{\nu}$ is known, one can estimate the parameters of a fOU on this dynamic~\cite{XiaoWangYu2018}. Given a time series $Y_1,...,Y_R$ following a fOU, we estimate the (constant) Hurst exponent as
$$\widehat H=\frac{1}{2}\log_2\left(\frac{M'_{R}(t;Y)}{M_{R}(t;Y)}\right).$$
We estimate the diffusion and mean-reversion parameters respectively by
\begin{equation}\nonumber
    \widehat{\eta} = \sqrt{\frac{1}{R(4-4^{\widehat H})}\sum\limits_{i=3}^{R}\left\vert Y_{i}-2Y_{i-1}+Y_{i-2}\right\vert^2}\quad \text{and} \quad \widehat{\lambda}=\left(\frac{R\sum\limits_{i=1}^{R}Y_{i}^2-\left(\sum\limits_{i=1}^{R}Y_{i}\right)^2}{R^2\widehat{\eta}^2\widehat{H}\Gamma(2\widehat{H})}\right)^{-1/2\widehat{H}}.
\end{equation}

\subsection{Forecasting procedure}\label{sec:forecast}

We consider a time series of log-price observed at high-frequency, that is intraday prices, $\{\log P_{j/r}\}_{j\in\llbracket 1,N\rrbracket}$, where $N$ indicates the whole amount of available data, $r=N/R$ the number of observations each day, and $R$ the number of days covered by the series. In the FSRM setting, this series represents an MPRE. We list below the steps to implement the forecast of future price increments in this framework. From the past sign of daily increment $\mathcal P_{t-1,t}=\sgn\left(\log P_t -\log P_{t-1}\right)$, we want to forecast the future sign of increment at a scale of $\tau$ days, $\mathcal P_{t,t+\tau}$. We will apply the following procedure to real data in Section~\ref{sec:empirical}. 
\begin{enumerate}
    \item For every day $i\in\llbracket 1,R\rrbracket$, we estimate a local Hurst exponent of log-prices in a window corresponding to the whole day, that is we apply the estimator defined in equation~\eqref{eq:estimH} to the intraday series $\log P_t$ with a window of size $r-1$. In this way, we get a series of daily estimated Hurst exponents $\{\widehat{H}^{r-1}_i\}_{i\in\llbracket 1,R\rrbracket}$. Two consecutive estimates are thus based on data which do not overlap, so that the series does not contain spurious serial dependence. 
    \item We halve the series $\{\widehat{H}^{r-1}_i\}_{i\in\llbracket 1,R\rrbracket}$, which is suppose to follow a fOU in our framework. On the first part $\{\widehat{H}^{r-1}_i\}_{i\in\llbracket 1,\frac{R}{2}\rrbracket}$ we estimate the parameters of the fOU, as described in Section~\ref{sec:estim}.
    \item Setting $\tau$ the number of future days for which the forecast is to be made, we apply Proposition~\ref{prop:1} on the second part of regularity series transformed with equation~\eqref{eq:norm_Ht}, $\{\widetilde{\widehat{H}^{r-1}_{i}}\}_{i\in\llbracket \frac{R}{2},R-\tau\rrbracket}$. Namely, we use the parameters $\widehat{H}$, $\widehat{\eta}$ and $\widehat{\lambda}$, estimated at the previous step, to compute the autocorrelation $\rho^H_{\tau\lambda}$, and we finally get the probability $\mathbb{P}(\widetilde{H}_{i+\tau}>1/2\vert \widetilde{\widehat{H}^{r-1}_{i}})$, for any day $i\in\llbracket \frac{R}{2},R-\tau\rrbracket$.
    \item Defining a threshold $\beta\in[1/2,1]$, we transform the above probability in one of the three following states: $+1$ if we have a strong presumption that $\widetilde{H}_{i+\tau}>1/2$, $-1$ if we have a strong presumption that $\widetilde{H}_{i+\tau}<1/2$, $0$ otherwise. We obtain this state by applying the function $f_{\beta}:x\mapsto \sgn(x-\beta)-\sgn(1-\beta-x)$ to $\mathbb{P}(\widetilde{H}_{i+\tau}>1/2\vert \widetilde{\widehat{H}^{r-1}_{i}})$.
    \item Finally, using the daily close price data $\{P_{i}\}_{i\in\llbracket\frac{R}{2}-1,R-\tau\rrbracket}$, we forecast the sign of the future increment of prices $\mathcal P_{i,i+\tau}$ with the following predictor: $\check{\mathcal P}^{\beta}_{i,i+\tau}=\mathcal P_{i-1,1}\times f_{\beta}\left(\mathbb{P}(\widetilde{H}_{i+\tau}>1/2\vert \widetilde{\widehat{H}^{r-1}_{i}})\right)$.
\end{enumerate}

The rationale of the predictor $\check{\mathcal P}^{\beta}_{i,i+\tau}$ is that if we predict a Hurst exponent above (respectively below) $1/2$ the future price increment will more probably have the same (resp. opposite) sign as the past one.

\subsection{Evaluation of the quality of the forecast}\label{sec:evalforecast}

The forecasting procedure, for a given threshold $\beta$, leads to the following hit rate
$$\hat{p}(\tau,\beta) = \frac{1}{n(\beta)-\tau}\sum\limits_{i=1}^{n(\beta)-\tau} \indic_{\check{\mathcal P}^{\beta}_{i,i+\tau}=\mathcal P_{i,i+\tau}},$$
where $n(\beta)=R(\beta)/2$ is the number of days in the second part of the series such that $\check{\mathcal P}^{\beta}_{i,i+\tau}\neq 0$. In order to determine whether the hit rate is significantly different from $1/2$, we apply a binomial test, for the null hypothesis being $\hat{p}(\tau,\beta)=1/2$ and the alternative $\hat{p}(\tau,\beta)>1/2$. The use of a binomial test is legitimate only if the hit variables $\indic_{\check{\mathcal P}^{\beta}_{i,i+\tau}=\mathcal P_{i,i+\tau}}$ are independent of each other. We will thus test this independence assumption of the hit variables with a BDS test, whose null hypothesis is the i.i.d. property~\cite{BDS1996}.

\section{Empirical study}\label{sec:empirical}

In this section, we evaluate the predictive performance of our method on three major stock indices at a 1-minute sampling: S\&P 500 index (SP500), Dow Jones index (DJI), and NYSE Composite index (NYSE). Table~\ref{tab:data} describes the three datasets. The analysis explores how the hit rate $\hat{p}(\tau,\beta)$, its statistical significance, and data utilization $n(\beta)-\tau$ vary as a function of the parameter $\beta$, which controls the filtering level applied to the data.

\begin{table}[!ht]
\centering
\begin{minipage}{0.7\textwidth}\caption{\small \label{tab:data} Overview of data used in the empirical analysis.}
\end{minipage}
\begin{tabular}{>{\raggedright}m{2.5cm} m{3.8cm} >{\raggedright}m{3cm} >{\raggedright}m{3cm} m{2cm}}
\toprule
\textbf{Data} &  \textbf{$N$ (Observations)} & \textbf{Starting date} & \textbf{Ending date} & \textbf{$R$ (Days)} \\
\midrule
\textbf{SP500} & 1,107,216 & 2010-03-29 09:30 & 2021-05-05 16:05 & 2,796 \\
%\vspace{.2cm}
\textbf{DJI} & 1,107,216 & 2010-03-29 09:30 & 2021-05-05 16:05 & 2,796 \\
%\midrule
%\vspace{.2cm}
\textbf{NYSE} & 1,474,308 & 2006-07-20 09:30 & 2021-05-05 16:05 & 3,723 \\
\bottomrule
\end{tabular}
\vspace{.2cm}
\end{table}

Applying the forecasting procedure exposed in Section~\ref{sec:forecast}, we estimate first daily Hurst exponents. Then, using a fOU specification for this series of Hurst exponents, we estimate the parameters $H$, $\lambda$, and $\eta$ of the fOU. Tables~\ref{tab:stima1} and~\ref{tab:stima2} show descriptive statistics respectively of the time series $\widehat H_t^{\eta}$ and of the estimates $\widehat H$, $\widehat \lambda$, and $\widehat \eta$, for each stock index.

\begin{table}[htbp]
\centering
\begin{minipage}{0.7\textwidth} \caption{\small \label{tab:stima1} Descriptive statistics of the estimated time-varying Hurst exponent $H_t$.}
\end{minipage}
\begin{tabular}{>{\raggedright}m{2.5cm} m{2cm} >{\raggedright}m{2cm} >{\raggedright}m{2cm} m{2cm}}
\toprule
\textbf{Data} &  \textbf{Mean} & \textbf{Std} & \textbf{Skewness} & \textbf{Kurtosis} \\
\midrule
\textbf{SP500} & $0.5104$ & $0.1219$ & $-0.5350$ & $2.8956$ \\
%\vspace{.2cm}
\textbf{DJI} & $0.4745$ & $0.0948$ & $-0.6140$ & $3.1183$ \\
%\midrule
%\vspace{.2cm}
\textbf{NYSE} & $0.4878$ & $0.2159$ & $-0.7315$ & $2.0802$ \\
\bottomrule
\end{tabular}
\vspace{.2cm}
\end{table}

\begin{table}[htbp]
\footnotesize
\centering
\begin{minipage}{0.7\textwidth}\caption{\small \label{tab:stima2} Parameter estimates for the dynamic of $H_t$ with their confidence intervals (CI).}
\end{minipage}
\begin{tabular}{>{\raggedright}m{2.5cm} m{3.5cm} m{3.5cm} m{3.5cm}}
\toprule
& $\widehat{H}$ (CI 95\%) & $\widehat{\eta}$ (CI 95\%) & $\widehat{\lambda}$ (CI 95\%) \\
\midrule
\textbf{SP500} & $0.0898 (0.0852, 0.0944)$ & $0.1049 (0.0997, 0.1101)$ & $0.0502 (0.0477, 0.0527)$ \\
\textbf{DJI} & $0.1851 (0.1758, 0.1944)$ & $0.0578 (0.0549, 0.0607)$ & $0.0990 (0.0941, 0.1040)$ \\
\textbf{NYSE} & $0.0763 (0.0725, 0.0801)$ & $0.1977 (0.1878, 0.2076)$ & $0.0049 (0.0047, 0.0051)$ \\
\bottomrule
\end{tabular}
\vspace{.2cm}
\end{table}

\begin{figure}[htbp]
    \centering
    \includegraphics[width=1\linewidth]{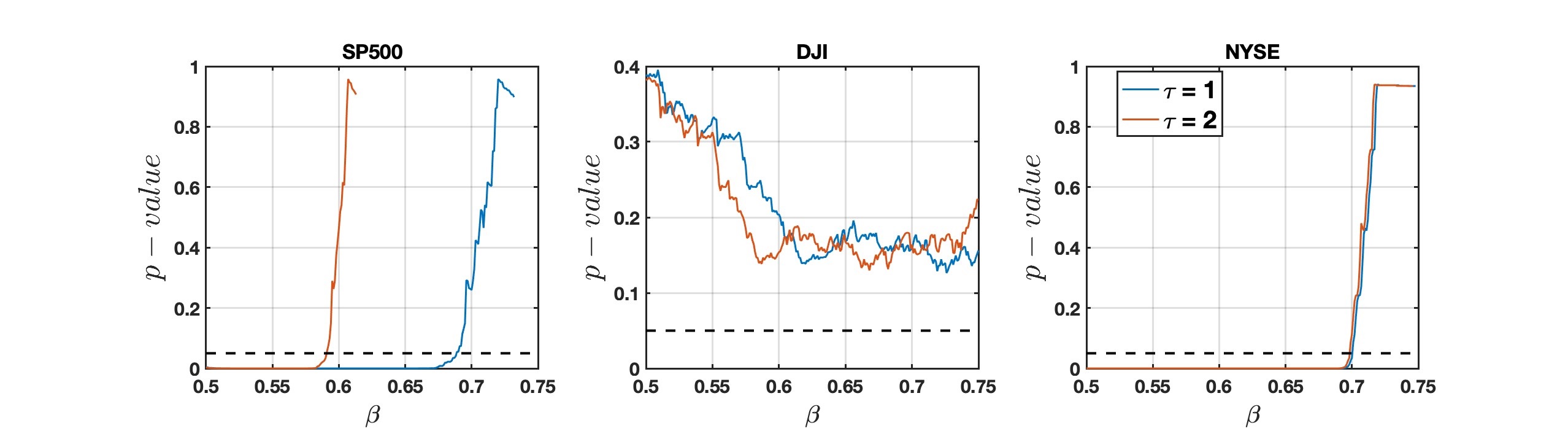}
    \begin{minipage}{0.7\textwidth}\caption{\small BDS test p-values as a function of the threshold parameter $\beta$ for SP500, DJI, and NYSE. Results are shown for prediction lags $\tau=1$ (blue) and $\tau=2$ (orange). The dashed horizontal line indicates the $5\%$ significance level. 
    \label{fig:independent}}
    \end{minipage}
\end{figure}

As explained in Section~\ref{sec:evalforecast}, the evaluation of the performance of the predictors is reliable only if the hit variables $\indic_{\check{\mathcal P}^{\beta}_{i,i+\tau}=\mathcal P_{i,i+\tau}}$ are independent of each other. We apply a BDS test, with dimension parameter equal to 3, for the hit variables corresponding to the three datasets, as displayed in Figure~\ref{fig:independent}. We observe that the p-values rise above the 0.05 threshold, for large values of $\beta$ in the case of SP500 and NYSE, and for all values of $\beta$ in the case of DJI, implying that the null hypothesis of independence cannot be rejected for the corresponding values of $\beta$. 

\begin{figure}[!ht]
    \centering
    \includegraphics[width=1\linewidth]{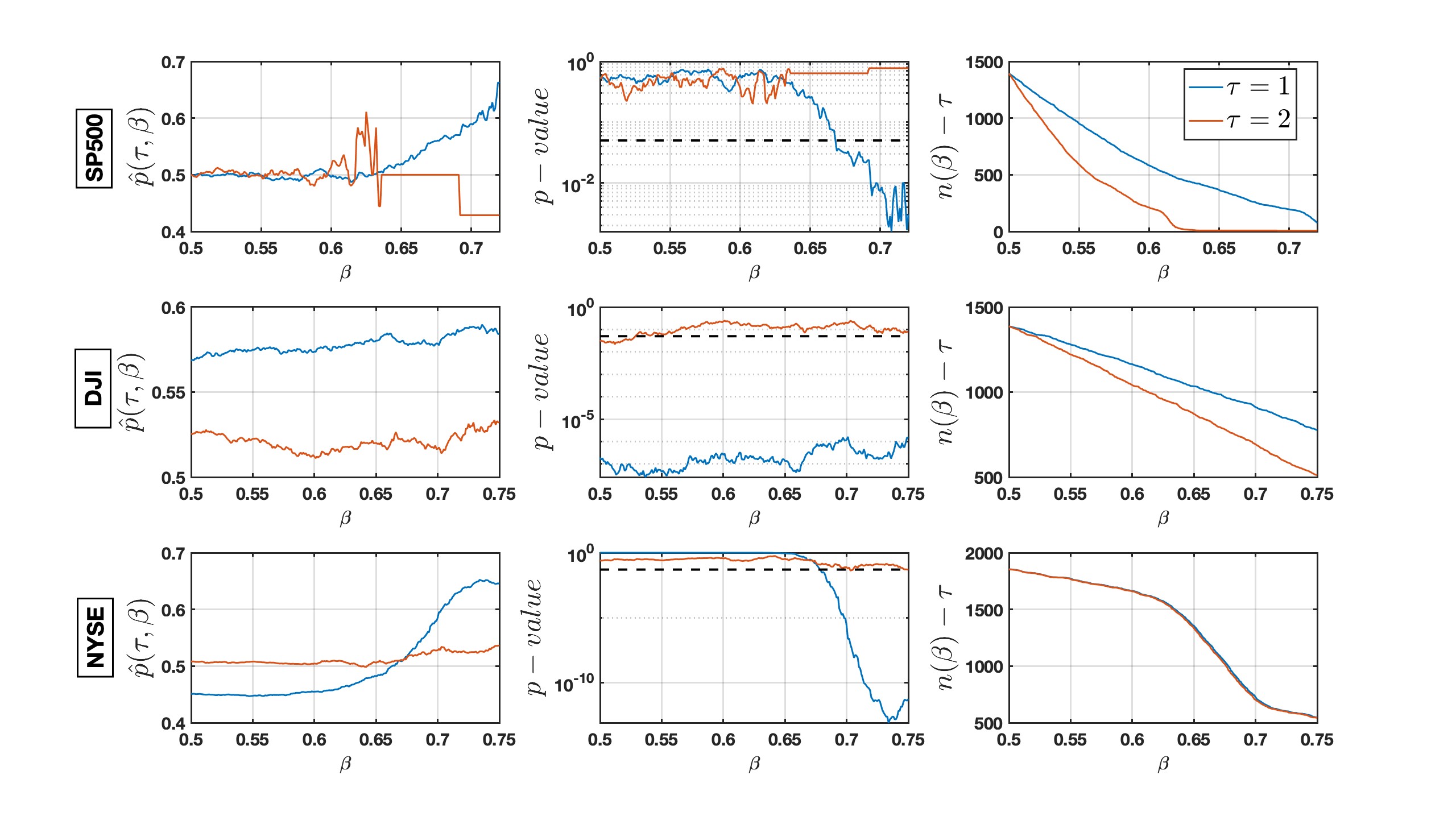}
    \begin{minipage}{0.7\textwidth}\caption{\small Predictive performance analysis for SP500, DJI, and NYSE as a function of the threshold parameter $\beta$, for prediction lags $\tau=1$ (blue) and $\tau=2$ (orange). The first column reports the hit rate $\hat{p}(\tau,\beta)$, the second column shows the p-values when testing the null hypothesis $\hat{p}(\tau,\beta)=0.5$ (with the dashed line marking the $5\%$ significance level), and the third column shows the effective sample size $n(\beta)-\tau$ used after applying the filtering induced by $\beta$.
    \label{fig:prediction_index}}
   \end{minipage}
\end{figure}

The BDS test has legitimated the binomial test for evaluating the significance of the hit rate of our predictor in the following cases: for the SP500, when $\beta>0.69$ and $\tau=1$ or when $\beta>0.59$ and $\tau=2$; for the DJI, whatever $\beta\in[0.5,0.75]$ and $\tau\in\{1,2\}$; for the NYSE, when $\beta>0.7$ and $\tau\in\{1,2\}$. Figure~\ref{fig:prediction_index} now presents a comprehensive view of how the hit rate $\hat{p}(\tau,\beta)$ behaves with respect to $\beta$, along with its statistical significance and the size of the filtered dataset. A p-value below 0.05 indicates a hit rate statistically different from $1/2$. The predictor for SP500 is significantly performing only when $\tau=1$ and for high value of $\beta$. We note that, in this case, we don't reject the independence of the hits only for $\beta>0.69$, what happens to approximatively 200 days out of 1400, with a hit rate of $60\%$. The predictor for DJI seems to be efficient for $\tau=1$ only, but whatever $\beta$, with a hit rate above $57\%$. For the NYSE dataset, the predictor is significantly performing only for $\tau=1$ and high value of $\beta$. Its is validated, because of the independence of the hits, for $\beta>0.7$, that is for approximatively 700 days out of 1850, with a hit rate above $60\%$. 

\section{Conclusion}\label{sec:conclusion}

Starting from the FSRM, a stochastic regularity model describing the dynamic of prices in a multifractal way, we have studied several properties of the fOU process that leads the dynamic of the regularity in the FSRM. We have thus provided the variance and the autocorrelation function of the fOU process and, more importantly in a financial perspective, its serial information. We have showed numerically that there are two different possible regimes for the fOU, depending on the value of the mean-reversion parameter, as for a delamperized fBm~\cite{BroutyGarcin2024}. When $H<1/2$, we have observed a non-zero, but very low information in the stationary regime, when the fOU is very different from an fBm. Empirical evidence confirms that the proposed prediction scheme identifies statistically significant patterns in equity markets, highlighting localized inefficiencies that are aligned with daily data frequency. This work leads to a better understanding of the fOU and of the FSRM and it finally opens the door to financial applications where the forecast of such a process matters, like in statistical arbitrage.

\section*{Acknowledgement}

MG acknowledges the support of the Chaire ``Deep Finance Statistics'' between QRT, Ecole Polytechnique and its foundation.

\bibliographystyle{plain}%unsrt}
\bibliography{Bibliography} 

\appendix

\section{Proof of Theorem~\ref{thm:autocorr_fOU}}\label{sec:thm_autocorr_fOU}

\begin{proof}
    Starting from equation~\eqref{eq:cov_fOU} in which we
    set $s=0$, we have
    \begin{equation}\nonumber
        \Var(Y_t^H) = \eta^2\frac{\Gamma(2H+1)\sin(\pi H)}{\pi\lambda^{2H}}\int_0^{\infty}\frac{x^{1-2H}}{1+x^2}dx.
    \end{equation}
    Defining the quantity $p=1-2H$, we can compute the integral $\int_0^{\infty}\frac{x^p}{1+x^2}dx$ using the residue theorem in the complex plane, where the integrand has two poles in $\pm i$, for $\vert p\vert<1$, i.e. for $0<H<1$. It holds
    \begin{equation}\nonumber
        \int_0^{\infty}\frac{x^p}{1+x^2}dx = \frac{2i\pi}{1-e^{2ip\pi}}\sum_{j = \pm i}\Res\Big(\frac{z^p}{1+z^2},j\Big) = \frac{\pi}{2\cos(p\pi/2)}.
    \end{equation}
    Finally the variance is
    \begin{equation}\nonumber
        \Var(Y_t^H) = \eta^2\frac{\Gamma(2H+1)\sin(\pi H)}{\pi\lambda^{2H}}\frac{\pi}{2\cos(\pi/2-\pi H)}=\frac{\eta^2\Gamma(2H+1)}{2\lambda^{2H}},
    \end{equation}
    while the autocorrelation function is
    \begin{equation}\nonumber
        \rho(Y_t^H,Y_{t+s}^H) = \frac{\Cov(Y_t^H,Y_{t+s}^H)}{\Var(Y_t^H)} = \frac{2\sin(\pi H)}{\pi}\int_0^{\infty}\cos(\lambda sx)\frac{x^{1-2H}}{1+x^2}dx.
    \end{equation}
 \end{proof}
 
\section{Proof of Theorem~\ref{thm:MI_FOU}}\label{sec:thm_MI_FOU}

\begin{proof}
    For any $t>0$, let $A = H_{tm+m}$ and $B = H_{tm}$. Knowing that $\sgn(\widetilde H_t-1/2)=\sgn(H_t-1/2)$, where $\sgn$ is the signum function, and thus that relations of the type $\mathbb{P}(\widetilde H_t>1/2\vert \widetilde H_{t-m}\leq 1/2) = \mathbb{P}(H_t>1/2\vert H_{t-m}\leq 1/2)$ hold, we can use the equations \eqref{eq:IM} and \eqref{eq:conditional_entropy} to write the serial information as:
\begin{align*}
    \textstyle
    & I^2_m = 1 - E(J_{m,.+1}\vert J_{m,.}) \\
    &    \quad\,=1 + \mathbb{P}(J_{m,.}=1)[f(\mathbb{P}(J_{m,.+1}=1\vert J_{m,.}=1))+f(\mathbb{P}(J_{m,.+1}=0\vert J_{m,.}=1))]\\
    &  \quad\quad\,+ \mathbb{P}(J_{m,.}=0)[f(\mathbb{P}(J_{m,.+1}=1\vert J_{m,.}=0))+f(\mathbb{P}(J_{m,.+1}=0\vert J_{m,.}=0))]\\
    &   \quad\, = 1 + \mathbb{P}(B>1/2)[f(\mathbb{P}(A>1/2\vert B>1/2))+f(\mathbb{P}(A\leq 1/2\vert B>1/2))]\\
    &   \quad\quad\,+ \mathbb{P}(B\leq 1/2)[f(\mathbb{P}(A>1/2\vert B\leq 1/2))+f(\mathbb{P}(A\leq 1/2\vert B\leq 1/2))].
\end{align*}
Noting that $\mathbb{P}(B>1/2)=\mathbb{P}(B\leq1/2)=1/2$ and that the events $A>1/2$ and $A\leq1/2$ are complementary, we get
\begin{equation}\label{eq:infoproof_fOU}
\begin{array}{ccl}
I^2_m & = & 1 + \frac{1}{2}\big[f(\mathbb{P}(A>1/2\vert B>1/2))+f(1-\mathbb{P}(A>1/2\vert B>1/2))\big]\\
& &+ \frac{1}{2}\big[f(\mathbb{P}(A>1/2\vert B\leq1/2))+f(1-\mathbb{P}(A>1/2\vert B\leq1/2))\big].
\end{array}\end{equation}
The vector $(A,B)'$ is Gaussian of mean $(1/2,1/2)'$ and, following Theorem~\ref{thm:autocorr_fOU}, of covariance matrix 
\begin{equation}\nonumber
    \Sigma_{AB} = \eta^2\frac{\Gamma(2H+1)}{2\lambda^{2H}}
    \begin{pmatrix}
        1 & \rho\\
        \rho & 1
    \end{pmatrix}.
\end{equation}
Some computations provide the determinant $\vert\Sigma_{AB}\vert = \eta^4\frac{\Gamma(2H+1)^2}{4\lambda^{4H}}(1-\rho^2)$ and 
\begin{equation}\nonumber
    \Sigma_{AB}^{-1} = \frac{2\lambda^{2H}}{\eta^2\Gamma(2H+1)(1-\rho^2)}
    \begin{pmatrix}
        1 & -\rho\\
        -\rho & 1
    \end{pmatrix}.
\end{equation}
We can calculate the joint probability as
\begin{equation}\label{eq:jointproba}
\begin{array}{cl}
%\textstyle
 & \mathbb{P}(A>1/2,B\leq 1/2) \\
= & \frac{1}{2\pi\vert\Sigma_{AB}\vert^{1/2}}\int_{-\infty}^{1/2}\int_{1/2}^{+\infty}\text{exp}\left(-\frac{1}{2}
    \begin{pmatrix}
        x - \frac{1}{2} & y - \frac{1}{2}
    \end{pmatrix}
    \Sigma_{AB}^{-1}
    \begin{pmatrix}
        x - \frac{1}{2}\\
        y - \frac{1}{2}
    \end{pmatrix}
    \right)dxdy \\
= & \frac{1}{2\pi\vert\Sigma_{AB}\vert^{1/2}}\int_{-\infty}^{0}{\int_{0}^{+\infty}{\text{exp}\left(-\frac{1}{2}\frac{2\lambda^{2H}}{\eta^2\Gamma(2H+1)(1-\rho^2)}
    \begin{pmatrix}
        x & y
    \end{pmatrix}
    \begin{pmatrix}
        1 & -\rho\\
        -\rho & 1
    \end{pmatrix}
    \begin{pmatrix}
        x\\
        y
    \end{pmatrix}
    \right)dx}dy}\\
 = & \frac{1}{2\pi\vert\Sigma_{AB}\vert^{1/2}}\int_{-\infty}^{0}{\left(\int_{0}^{+\infty}\text{exp}\Big(-\frac{1}{2}\frac{2\lambda^{2H}(x-\rho y)^2}{\eta^2\Gamma(2H+1)(1-\rho^2)}
    \Big)dx\right)\exp\left(-\frac{1}{2}\frac{2\lambda^{2H}y^2}{\eta^2\Gamma(2H+1)}\right)dy}.
\end{array}\end{equation}
Using the substitutions $\omega = \sqrt{\frac{2}{\Gamma(2H+1)(1-\rho^2)}}\frac{\lambda^H}{\eta}(x-\rho y)$ and $z=-\sqrt{\frac{2}{\Gamma(2H+1)}}\frac{\lambda^H}{\eta}y$, we get 
$$\begin{array}{ccl}
 \mathbb{P}(A>1/2,B\leq 1/2) & = & \int_0^{+\infty}\left(\int_{\frac{\rho z}{\sqrt{1-\rho^2}}}^{+\infty}\frac{e^{-\frac{\omega^2}{2}}}{\sqrt{2\pi}}d\omega\right)\frac{e^{-\frac{z^2}{2}}}{\sqrt{2\pi}}dz \\
& = & \int_{0}^{+\infty}N\left(-\frac{\rho z}{\sqrt{1-\rho^2}}\right)g(z)dz \\
& = & \frac{1}{4}-\frac{1}{2\pi}\arctan\Bigg(\frac{\rho}{\sqrt{1-\rho^2}}\Bigg),
\end{array}$$
where we used in the last step a known result on the integral of a product of the standard Gaussian pdf $g$ and cdf $N$~\cite[Lemma 1]{Garcin2022}. Noting that $\mathbb{P}(B\leq 1/2)=1/2$, we get:
\begin{equation}\label{eq:probacond1}
\mathbb{P}(A>1/2\vert B\leq 1/2) = \frac{1}{2}-\frac{1}{\pi}\arctan\left(\frac{\rho}{\sqrt{1-\rho^2}}\right).
\end{equation}
Similarly, we also obtain
\begin{equation}\label{eq:probacond2}
\mathbb{P}(A>1/2\vert B> 1/2) = \frac{1}{2}+\frac{1}{\pi}\arctan\left(\frac{\rho}{\sqrt{1-\rho^2}}\right).
\end{equation}
Finally, using equations~\eqref{eq:infoproof_fOU}, \eqref{eq:probacond1}, and~\eqref{eq:probacond2}, we get the result displayed in the theorem:
\begin{equation}\nonumber
\textstyle
    I^2_m = 1 + f\left(\frac{1}{2}-\frac{1}{\pi}\arctan\left(\frac{\rho}{\sqrt{1-\rho^2}}\right)\right) + f\left(\frac{1}{2}+\frac{1}{\pi}\arctan\left(\frac{\rho}{\sqrt{1-\rho^2}}\right)\right).
\end{equation}
\end{proof}
 
\section{Proof of Proposition~\ref{prop:1}}\label{sec:prop_1}

\begin{proof}
The stationary fOU process is distributed like $(H_{mi},H_{m(i+1)})'\sim N((1/2,1/2)',\Sigma)$, where $\Sigma= \theta^2\begin{pmatrix} 
    1 & \rho^H_{m\lambda} \\ \rho^H_{m\lambda} & 1
\end{pmatrix} $ and
$\theta^2=\eta^2\Gamma(2H+1)/2\lambda^{2H}$, after Theorem~\ref{thm:autocorr_fOU}. We also note that the determinant is $|\Sigma|=\theta^4(1-(\rho^H_{m\lambda})^2)$. Therefore, using the joint probability provided in equation~\eqref{eq:jointproba} and noting $\check y=y-1/2$ and $\check x=x-1/2$, the conditional density $f_{H_{m(i+1)}|H_{mi}}$ follows
\begin{align*}
f_{H_{m(i+1)}|H_{mi}}(y \vert x) & = \frac{f_{H_{m(i+1)},H_{mi}}\left(\frac{1}{2}+\check y,\frac{1}{2}+\check x\right)}{f_{H_{mi}}\left(\frac{1}{2}+\check x\right)} \\
 & =  \frac{\frac{1}{2\pi|\Sigma|^{1/2}}\exp\left(-\frac{1}{2}\frac{(\check y-\rho^H_{m\lambda} \check x)^2}{\theta^2(1-(\rho^H_{m\lambda})^2)}
    \right)\exp\left(-\frac{1}{2}\frac{\check x^2}{\theta^2}\right)}{\frac{1}{\sqrt{2\pi\theta^2}}\exp\left(-\frac{1}{2}\frac{\check x^2}{\theta^2}\right)} \\
& =\frac{1}{\sqrt{2\pi\theta^2(1-(\rho^H_{m\lambda})^2)}}\exp\left(-\frac{1}{2}\frac{(\check y-\rho^H_{m\lambda} \check x)^2}{\theta^2(1-(\rho^H_{m\lambda})^2)}
    \right).
\end{align*}
The conditional density $f_{H_{m(i+1)}|H_{mi}}\left(\frac{1}{2}+\check y\vert\frac{1}{2}+\check x\right)$ is thus a Gaussian density in $\check y$, of mean $\rho^H_{m\lambda} \check x$ and variance $\theta^2(1-(\rho^H_{m\lambda})^2)$. A simple substitution thus provides us with the conditional probability
\begin{align*}
p(1\vert x) & =\int_0^{\infty}f_{H_{m(i+1)}|H_{mi}}\left(\left.\frac{1}{2}+\check y\right| x\right) d\check y \\
 & = \int_0^{\infty} g_{\rho^H_{m\lambda}(x-1/2),\theta^2(1-(\rho^H_{m\lambda})^2)}(\check y)d\check y,
\end{align*}
where $g_{a,b^2}$ is the Gaussian density of mean $a$ and variance $b^2$. Noting that a substitution $z=(y-a)/|b|$ leads to $\int_0^{\infty}g_{a,b^2}(y)dy=\int_{-a/|b|}^{\infty}g_{0,1}(z)dz=N(a/|b|)$, we finally get
$$p(1\vert x)=N\left(\frac{\rho^H_{m\lambda}(x-1/2)}{|\theta|\sqrt{1-(\rho^H_{m\lambda})^2}}\right).$$
\end{proof}

\end{document}